\documentclass[Review,sagev,times,doublespace]{sagej}

\usepackage{moreverb,url}

\usepackage[colorlinks,bookmarksopen,bookmarksnumbered,citecolor=black,urlcolor=black]{hyperref}

\usepackage{natbib}
\usepackage{subfigure}
\usepackage{graphicx} 
\usepackage{float}
\usepackage{bm} 
\usepackage{array}
\usepackage{amsmath,amssymb}
\usepackage{booktabs}
\usepackage{multirow}

\newtheorem{theorem}{Theorem}
\newtheorem{remark}{Remark}

\newcommand\BibTeX{{\rmfamily B\kern-.05em \textsc{i\kern-.025em b}\kern-.08em
T\kern-.1667em\lower.7ex\hbox{E}\kern-.125emX}}

\def\volumeyear{2016}

\begin{document}

\runninghead{Li et al.}

\title{Multicategory Matched Learning for Estimating Optimal Individualized Treatment Rules in Observational Studies with Application to a Hepatocellular Carcinoma Study}

\author{Xuqiao Li\affilnum{1}, Qiuyan Zhou\affilnum{1}, Ying Wu\affilnum{2} and Ying Yan\affilnum{1}}

\affiliation{\affilnum{1}School of Mathematics, Sun Yat-sen University, Guangzhou, China\\
\affilnum{2}School of Statistics and Data Science, LPMC and KLMDASR, Nankai University, Tianjin, China}

\corrauth{Ying Yan, School of Mathematics, Sun Yat-sen University, Guangzhou, China.\\
Email: \href{mailto:yanying7@mail.sysu.edu.cn}{yanying7@mail.sysu.edu.cn} \\
 Ying Wu, School of Statistics and Data Science, LPMC and KLMDASR, Nankai University, Tianjin, China.\\
Email: \href{mailto:ywu@nankai.edu.cn}{ ywu@nankai.edu.cn}}

\begin{abstract}
One primary goal of precision medicine is to estimate the  individualized treatment rules (ITRs) that optimize patients' health outcomes based on individual characteristics. Health studies with multiple treatments  are commonly seen in practice. However, most existing ITR estimation methods   were developed for the studies with binary treatments. Many require that the outcomes are fully observed. In this paper, we propose a  matching-based machine learning method to estimate  the optimal ITRs in  observational studies with multiple treatments when the outcomes are fully observed or  right-censored. We establish  theoretical property for the proposed method. It is compared with the existing %inverse probability weighting-based ITR estimation
competitive methods  in  simulation studies and a  hepatocellular carcinoma study.
\end{abstract}


\maketitle

\section{Introduction}
\label{sec1}
In recent years, precision medicine has drawn great attention in medical research and practice.\citep{collins2015new,schork2015personalized} A primary goal of  precision medicine is to  maximize patients' health outcomes by tailoring treatment strategies for each patient based on  individualized characteristics.\citep{kosorok2019precision} Individualized treatment rules formalize this idea, 
and the optimal ITR is a decision rule based on  patients' characteristics to choose an appropriate treatment that maximizes the expected health outcomes. A practical  guideline on personalized treatment recommendation for a specific health care problem  can then be formulated accordingly.

Statistical methods for estimating  optimal ITRs with two treatments were extensively studied in the literature. These methods can be roughly divided into regression-based methods and classification-based methods (see e.g. Blatt et al.,\cite{blatt2004learning} Qian and Murphy,\cite{qian2011performance} Zhao et al.,\cite{zhao2012estimating} Liu et al.,\cite{liu2018augmented} Zhao et al.,\cite{zhao2019efficient} and Wu et al.\cite{wu2020matched}). For example, \textcolor{black}{Q-learning \citep{watkins1992q,qian2011performance} fits the conditional mean of the outcome given treatment and covariates and then determines the optimal ITRs. A-learning \citep{murphy2003optimal,blatt2004learning,robins2004optimal} models the contrast of treatment effects. Q-learning, A-learning, and several subsequent works \citep{lu2013variable,laber2014interactive,song2015penalized,shi2018high} are regression-based methods. Nevertheless, their  performances critically rely on  correct model specification.} Alternatively,  classification-based methods directly search for the optimal ITRs. Outcome-weighted learning (O-learning),\citep{zhao2012estimating} among others, recasts the ITR 
problem  into a weighted classification problem, which is  solved by machine learning algorithms including weighted support vector machine (SVM). Augmented O-learning \citep{liu2018augmented}  accommodates negative weights and shows improved performance compared to O-learning. O-learning for the  ordinal treatment settings was developed.\citep{chen2018estimating} \textcolor{black}{Moreover, approaches utilizing augmented inverse probability weighting  have been proposed, which combine  the advantages of the aforementioned regression  and outcome-weighted  strategies.\cite{zhang2012robust,zhang2013robust}} Some research efforts have been made in recent years for  estimating optimal ITRs using  survival data  (see e.g. Zhao et al.,\cite{zhao2015doubly} Cui et al.,\cite{cui2017tree} Leete et al.,\cite{leete2019balanced} and Xue et al.\cite{xue2021multicategory}).

In the scenario of multiple treatments, a direct extension of the aforementioned methods using multiple pairwise  comparisons can be problematic and suboptimal.\citep{XuWood2020}  Recently, several methods were developed for  learning optimal ITRs directly  with  multiple treatments. For example,
Lou et al.\cite{lou2018optimal} extended O-learning to accommodate multiple treatments using  the vector hinge loss function, assuming that the  propensity score is known. Huang et al.\cite{huang2019multicategory} proposed a doubly robust version of O-learning with generic surrogate loss functions for  randomized controlled trials and observational studies. Qi et al.\cite{qi2020multi} proposed an insightful angle-based direct learning method  with simple and good interpretations. Zhang et al.\cite{Zhang2020sinica} integrated  angle-based  learning and O-learning. Xue et al. \cite{xue2021multicategory} extended angle-based  learning to the survival setting.

In observational studies, existing ITR estimation methods 
usually adjust for confounding bias via inverse probability weighting (IPW), where the propensity score is estimated by logistic regression or machine learning.\citep{lee2010improving} Potential drawbacks of the IPW-based methods for estimating optimal ITRs  include  high variability  due to extreme weights and inferior performance due to model misspecification.  Moreover, when the number of treatment options increases, the IPW-based methods may perform worse because the issue of extreme weights becomes  more prevalent. 

Recently, Wu et al.\cite{wu2020matched} proposed matched learning (M-learning) for observational studies with binary treatments, and showed that   O-learning  is a special case of M-learning. M-learning uses matching, not IPW, to adjust for confounding bias in observational studies. \textcolor{black}{Wu et al. \cite{wu2020matched} demonstrated the competitive performance of M-learning. Particularly, compared with Q-learning, the superiority of M-learning becomes more pronounced when the outcome  model is misspecified.  M-learning greatly outperforms O-learning when the propensity score model is incorrect.}  Nevertheless, M-learning only deals with binary treatments. A direct extension of M-learning to the multicategory situation using  multiple pairwise  comparisons is undesirable.

In this paper, we  extend  M-learning  to the setting of multiple treatments in observational studies, named multicategory M-learning.  
We  reformulate the matching-based value function in Wu et al.\cite{wu2020matched} from the imputation viewpoint. This viewpoint allows us to circumvent  pairwise comparisons in M-learning. We then   generalize  the reformulated value function to the setting with multiple treatment arms, where  we adapt multicategory matching (see e.g. Fr\"olich\cite{frolich2004programme} and Yang et al.\cite{yang2016propensity}).  The optimal ITR is obtained by maximizing the multicategory value function. We show that the maximization problem is equivalent to a multicategory weighted misclassification problem, which can be solved using {multicategory  SVM algorithms (see e.g. Lee et al.,\cite{lee2004multicategory} Liu and Yuan,\cite{liu2011reinforced} Zhang and Liu,\cite{zhang2014angle} and Zhang et al.\cite{zhang2016reinforced}). Among them, Zhang et al.\cite{zhang2016reinforced} proposed the reinforced angle-based multicategory SVM (RAMSVM) method  that enjoys fast computational speed and the Fisher consistency property. We adopt RAMSVM to solve the multicategory weighted classification problem for the proposed multicategory M-learning method  in this paper.}

We establish   theoretical property  for the proposed  method under suitable assumptions.  Furthermore, we extend this method to accommodate right-censored survival data. Motivated by the imputation idea in Cui et al.,\cite{cui2017tree}  censored observations are imputed nonparametrically by random survival forest.\citep{ishwaran2008random}

This work is partly motivated by an observational study of   hepatocellular carcinoma  (HCC), a common type of liver cancer.  The data were collected between July 2009 and May 2019 in Tianjin, China.\citep{zhang2020EASL,zhang2020gut} The data  pose challenges from several aspects. First, there were three types of treatments for HCC patients in this study, including liver transplantation (LT), liver resection (LR), and  local ablation (LA). Second, the clinical outcomes  were right-censored. Moreover, confounding bias needs to be adjusted using observed individual-level information. M-learning is an appealing   learning method to identify optimal ITRs in observational studies, but it is only intended for  continuous, ordinal or discrete outcomes with two treatments.\citep{wu2020matched}  The issues of multiple treatments and right-censoring are not tackled in M-learning. The proposed multicategory M-learning method in this paper  is   capable of dealing with right-censored survival data with multiple treatments, and thus is suitable for the HCC  study.

The rest of the paper is organized as follows. In Section \ref{sec2}, we introduce the basic framework  and M-learning. In Section \ref{sec3}, we present the proposed methodology and discuss the survival setting. Simulation
studies are presented in Section \ref{sec4}. Analysis of the HCC data is given in Section \ref{sec5}.
The article concludes with  discussions  in Section \ref{sec6}.  The Supplemental Material  includes the theoretical property of the proposed method and additional simulations.

\section{Formulation and Preliminary}\label{sec2}
We first formulate the  problem of estimating optimal ITRs with two treatments. Let $\bm{X} \in \mathcal{X}$ denote a patient's pretreatment $p$-dimensional covariates, where $\mathcal{X}$ is the covariate space. Let $A$ be the treatment  indicator taking values in $\{-1,1\}$. Let $R$ be the clinical outcome  assuming that  a larger value is more beneficial. The observed data are $n$ independent and identically distributed (i.i.d.) copies of $(\bm{X},A,R)$, denoted as $\{(\bm{X}_i,A_i,R_i), i=1,\cdots, n\}$.

An ITR $\mathcal{D}(\cdot)$ is a decision rule mapping from $\mathcal{X}$ to  $\{-1,1\}$. The central task is to estimate the optimal ITR that maximizes the expected clinical outcome in the patient population if this ITR were implemented. Let $V(\mathcal{D})=E^{\mathcal{D}}(R)$  be  the value function associated with the ITR $\mathcal{D}(\cdot)$. It is the expectation of $R$  given  $A=\mathcal{D}(\bm{X})$, i.e.,   the expected clinical outcome that follows  $\mathcal{D}(\cdot)$ to assign treatments.

Zhao et al.\cite{zhao2012estimating} re-expressed the value function to be 
\begin{equation}{\label{valuefun}}
	V(\mathcal{D})=E\left[\frac{ I\{A=\mathcal{D}(\bm{X})\}R} { \pi(A , \bm{X})}\right],
\end{equation}
where  $\pi(a,\bm{x}) = Pr(A=a|\bm{X}=\bm{x})$  is the propensity score, and then  proposed O-learning for  randomized controlled trials. O-learning searches for an optimal ITR, named $\mathcal{D}^*(\cdot)$, that maximizes the empirical value function:
$$\mathcal{D}^{*}=\underset{\mathcal{D}}{\arg \max } \frac{1}{n} \sum_{i=1}^{n} \frac{I\left\{A_{i}=\mathcal{D}\left(\bm{X}_{i}\right)\right\} R_{i}}{\pi\left(A_{i}, \bm{X}_{i}\right)}.$$

In  observational studies,  the propensity score $\pi(a , \bm{x})$ is unknown. It can be  estimated by logistic regression or  machine learning (see e.g. Lee et al.\cite{lee2010improving}), but extreme weights may exist   and  the  logistic model may be misspecified.   Alternatively, Wu et al.\cite{wu2020matched} proposed  M-learning based on the following matching-based value function:
\begin{equation}{\label{valuefun_mlearning}}
	\begin{aligned}
		\tilde{V}_{n}(\mathcal{D};g)
		=& \frac{1}{n} \sum_{i=1}^{n} \frac{1}{\left|\mathcal{M}_{i}\right|} \sum_{j \in \mathcal{M}_{i}}\{I\left(R_{j} \geq R_{i}, \mathcal{D}\left(\bm{X}_{i}\right)=-A_{i}\right)
		\\ + & I\left(R_{j} \leq R_{i}, \mathcal{D}\left(\bm{X}_{i}\right)=A_{i}\right) \}
		 \times g\left(\left|R_{j}-R_{i}\right|\right),
	\end{aligned}
\end{equation} 
{\color{black} where $\mathcal{M}_i$ denotes a matched set for subject $i$ that consists of subjects with similar covariate values as subject $i$ but received the opposite treatment $A_j=-A_i$.  Here,  similarity is determined by some sensible distance metric. We provide more discussions in Section \ref{sec3.2}. Therefore, subject $i$ and any subject in $\mathcal{M}_i$ formulates a matched pair with different treatments. The value function consists of $n\sum_{i=1}^n|\mathcal{M}_{i}|$ matched pairs. Within each  pair, the subject with  larger clinical outcome is more likely to have received the optimal treatment. The weighting function $g(\cdot)$ is monotonically increasing as the absolute outcome difference in the matched pair becomes larger.} 

The problem of searching for an optimal ITR by maximizing the value function \eqref{valuefun_mlearning} is  expressed as a weighted SVM problem with  the zero-one loss replaced by the hinge loss,  which can then be solved by existing quadratic programming software. {\color{black} In Section \ref{sec3.3}, we describe the algorithmic details.}

Before introducing the proposed methodology for multiple treatments, we reformulate the matching-based value function \eqref{valuefun_mlearning} from an imputation viewpoint. The pretreatment characteristics of subject $i$ and  the subjects in the matched set $\mathcal{M}_i$ are similar, and thus the  outcomes $\{R_j:\ j\in \mathcal{M}_i\}$ can be roughly  regarded as multiple  random draws from the distribution of subject $i$'s outcome if treatment $-A_i$ were assigned instead. In other words, the matched set $\mathcal{M}_i$ contains multiple imputed outcomes for  subject $i$.

For any $j\in {\mathcal{M}_i}$,  let
\begin{equation*}\label{matching_estimator}
	\widehat{R}^{(j)}_{i}(w) = \left\{ \begin{array}{l}
		{R_i}, \text{ if } w = {A_i} ,\\
		{R_j}, \text{ if } w =  - {A_i}.
	\end{array} \right.
\end{equation*}
 Noting that
\begin{equation*}
	\begin{aligned}
	&I\left(R_{j} \geq R_{i}, \mathcal{D}\left(\bm{X}_{i}\right)=-A_{i}\right)  + I\left(R_{j} \leq R_{i}, \mathcal{D}\left(\bm{X}_{i}\right)=A_{i}\right) \\
	 =& I\{\mathcal{D}(\bm{X}_i)=\underset{w \in \{-1,1\}}{\arg \max }  \widehat{R}^{(j)}_{i}(w)\},
	\end{aligned}
\end{equation*}
the  value function \eqref{valuefun_mlearning} is equivalent to
\begin{equation}{\label{valuefun_potential}}
	V_{n}(\mathcal{D} ; g)= n^{-1} \sum_{i=1}^{n}\left|\mathcal{M}_{i}\right|^{-1}  \sum_{j \in \mathcal{M}_{i}}  I\{\mathcal{D}(\bm{X}_i)=\underset{w \in \{-1,1\}}{\arg \max } \widehat{R}^{(j)}_{i}(w)\}  g\left(\left|R_{j}-R_{i}\right|\right).
\end{equation}
{\color{black} M-learning seeks to identify the decision rule that assigns the treatment associated with the larger outcome to each patient. This formulation avoids multiple pairwise comparisons and generalizes M-learning to multiple treatments straightforwardly.}

\section{Proposed Methodology}\label{sec3}
\subsection{Multicategory  Matched Learning}\label{sec3.2}
Now, we consider the general setting that there are $k\geq 3$ treatments in the observational study. The treatment space is $\mathcal{A}=\{1,2,\ldots,k\}$. 
An ITR $\mathcal{D}(\cdot)$ is a decision rule mapping from the covariate space $\mathcal{X}$ to  the treatment space  $\mathcal{A}$.  The observed data are  $\{(\bm{X}_i,A_i,R_i), i=1,\cdots, n\}$.
Let $\pi(a,\bm{x})$ denote the generalized propensity score $Pr(A=a|\bm{X}=\bm{x})$, $a=1,\cdots, k$.  

In the presence of  $k\geq 3$  treatment arms, a direct extension of  the original  value function \eqref{valuefun_mlearning}  is undesirable, because it entails multiple pairwise  comparisons within the matched pairs and thus can be suboptimal.\citep{XuWood2020} We instead generalize the alternative expression  \eqref{valuefun_potential} to tackle multiple treatments.

For each subject $i=1,\cdots, n$, we construct $k-1$ matched sets such that each set contains subjects with similar covariates  as subject $i$ but with treatment $w\neq A_i$. That is, for any treatment option $w\in \mathcal{A}-\{A_i\}$, we build the following matched set using covariate matching with replacement for subject $i$:
\begin{equation}\label{multiple_matchedset}
	\mathcal{M}^{(w)}_i=\{j : A_j=w, \ \ d(\bm{X}_{j},\bm{X}_i) \leq \delta^{(w)}_i\},
\end{equation}
where $d(\cdot,\cdot)$ denotes a distance measure in the  covariate space. The   Mahalanobis distance is a  common choice for $d(\cdot,\cdot)$ in practice. 
{\color{black} The threshold parameter $\delta^{(w)}_i$  is pre-specified and  non-negative, and the size of matched set may be different across subjects. In nearest neighbour matching, $\delta^{(w)}_i$ is the minimal distance between subject $i$ and the subjects that received treatment $w$.  In general, a well-specified threshold may increase the quality of matches.\citep{stuart2010matching}}  The subjects in the matched set $\mathcal{M}^{(w)}_i$ are similar to subject $i$ in the sense that they share similar pretreatment features.  The outcomes $\{R_j, \ j\in\mathcal{M}^{(w)}_i\}$ are  regarded as multiple random draws from the   distribution of subject $i$'s outcome if treatment $w$ were assigned instead of the actual treatment $A_i$.  For convenience of notation,   for  $w=A_i$, let $\mathcal{M}^{(A_i)}_i=\{i\}$ be the matched set consisting of only subject $i$. Then, the matched set $\mathcal{M}^{(w)}_i$ is defined for  each treatment arm $w=1,\cdots, k$.

For  treatment $w=1,\cdots, k$, we   select an arbitrary subject  from   $\mathcal{M}^{(w)}_i$, say subject $j_w$,  to construct a matched pair $(i,j_w)$.   Define $\boldsymbol{j}=(j_1,j_2,\cdots, j_k)$ to be  a matched tuple for subject $i$. Given subject $i$ and  the associated matched tuple $\boldsymbol{j}$, define  for  $w=1,\cdots, k$  that
\begin{equation*}\label{matching_estimator_multiple}
	\widehat{R}^{(\boldsymbol{j})}_{i}(w) = \left\{ \begin{array}{l}
		{R_i}, \text{ if } w = {A_i} ,\\
		{R_{j_w}}. \text{ if } w \neq {A_i}.
	\end{array} \right.
\end{equation*}
Hence $\widehat{R}^{(\boldsymbol{j})}_{i}(w) =R_{j_w}$ for any $w$ by noting that  ${j}_{A_i}=i$. The set 
$\{\widehat{R}^{(\boldsymbol{j})}_{i}(1),\cdots, \widehat{R}^{(\boldsymbol{j})}_{i}(k)\}=\{R_{j_1},\cdots, R_{j_k}\}$ contains the responses in the  matched tuple $\boldsymbol{j}$. They can be  regarded as   subject $i$'s  imputed outcomes if each of the $k$ treatments  were assigned. Let $\mathcal{M}_i=\mathcal{M}^{(1)}_i\times \cdots \times \mathcal{M}^{(k)}_i$ be the Cartesian product of the $k$ matched sets:
$\mathcal{M}_i=\{\boldsymbol{j}=(j_1,j_2,\cdots, j_k): j_l\in \mathcal{M}_i^{(l)}, l=1,\cdots, k \}$.
Therefore,  $\mathcal{M}_i$ contains  $\prod_{w\neq A_i}^k |\mathcal{M}^{(w)}_i|$ permutations to pick a  matched tuple $\boldsymbol{j}$ for subject $i$.

Now, we define the  multicategory matching-based value function  for any ITR $\mathcal{D}$:
\begin{equation}{\label{valuefun_multi}}
	\begin{aligned}
	V_n(\mathcal{D};g)= & \frac{1}{n} \sum_{i=1}^n \frac{1}{\prod\limits_{w\neq A_i} |\mathcal{M}^{(w)}_i|} \underset{\boldsymbol{j}  \in \mathcal{M}_i}{\sum}  I\{\mathcal{D}(\bm{X}_i)=\underset{w \in \{1,\ldots,k\}}{\arg\max} \widehat{R}^{(\boldsymbol{j})}_{i}(w)\} \\
	\times & g(R_{j_1},\cdots, R_{j_k}),
	\end{aligned}
\end{equation}
where $g(\cdot)$ is a  non-negative weighting function that encodes information of outcome differences.  The motivation is analogous to the imputation viewpoint   in Section \ref{sec2}. For any matched tuple $\boldsymbol{j}  \in \mathcal{M}_i$, we expect that the subject with  largest outcome has received the optimal treatment among the $k$ 
treatment options, and thus the optimal ITR should assign this treatment to subject $i$.  Multicategory M-learning maximizes $V_n(\mathcal{D};g)$ in \eqref{valuefun_multi} to obtain the optimal ITR. When $k=2$,  the multicategory value function $V_n(\mathcal{D};g)$ reduces to  the value function \eqref{valuefun_mlearning}, and  multicategory M-learning reduces to the original M-learning with binary treatments.\citep{wu2020matched}

Next, we discuss the choices for $g(\cdot)$. The simplest  weighting function is $g(R_{j_1},\cdots, R_{j_k})\equiv 1$.  If the contrast between the largest  outcome and the remaining outcomes is greater, then the likelihood is expected to be greater. It motivates us to 
consider $g(\cdot)$ to be the sum of the differences between the largest  outcome and the rest  outcomes:
\begin{equation}{\label{gweight1}}
	g_1(R_{j_1},\cdots, R_{j_k})=\sum\limits_{w = 1}^k (\max_{l\in \mathcal{A}} R_{j_l} -R_{j_w}),
\end{equation}
which denotes the total  reward for  subject $i$ if  the  $k-1$ non-optimal  treatments were replaced by the optimal one. We also consider $g(\cdot)$ to be the difference between the largest and  smallest  outcomes:
\begin{equation}{\label{gweight2}}
	g_2(R_{j_1},\cdots, R_{j_k})= \max_{l\in \mathcal{A}} R_{j_l} - \min_{l\in \mathcal{A}} R_{j_l},
\end{equation}
which is the maximum reward for  subject $i$  if the least favorable treatment were replaced by the optimal one. 

The weighting functions $g_1(\cdot)$ and $g_2(\cdot)$ are exchangeable.  Exchangeability is required to establish  theoretical property for the proposed method. Note that for the binary treatment setting with $k=2$,  $g_1(\cdot)$ and $g_2(\cdot)$ reduce to  $g(R_i, R_j)=|R_i-R_j|$ in  M-learning.\citep{wu2020matched}  

\subsection{Estimation Procedure}\label{sec3.3}
The proposed multicategory M-learning method maximizes $V_n(\mathcal{D};g)$ in \eqref{valuefun_multi}, or equivalently, minimizes 
\begin{equation}{\label{classification}}
	\sum_{i=1}^n  \underset{\boldsymbol{j}  \in \mathcal{M}_i}{\sum} W^{(\boldsymbol{j})}_i I(\mathcal{D}(\bm{X}_i)\neq Y_i^{(\boldsymbol{j})})
\end{equation}
to identify the optimal ITR, where  $W^{(\boldsymbol{j})}_i= n^{-1}\prod_{w\neq A_i} |\mathcal{M}^{(w)}_i|^{-1}g(R_{j_1},\cdots, R_{j_k})$ is the individual weight and $Y_i^{(\boldsymbol{j})}=\arg\max_{w \in \{1,\ldots,k\}} \widehat{R}^{(\boldsymbol{j})}_{i}(w)$ is the optimal treatment for subject $i$ given the matched tuple $\boldsymbol{j}$. This is essentially a multicategory weighted classification problem, where there are $n \prod_{w\neq A_i} |\mathcal{M}^{(w)}_i|$ training data points: 
$\{(\bm{X}_i, Y_i^{(\boldsymbol{j})}),\ \ i=1,\cdots, n,\ \ \boldsymbol{j}  \in \mathcal{M}_i\}$. Here, $\bm{X}_i$ is the feature and $Y_i^{(\boldsymbol{j})}$ is the  label taking values in $\{1,\cdots, k\}$. Therefore, the expression in \eqref{classification} is  a weighted sum of misclassification errors given any ITR $\mathcal{D}(\cdot)$.

It is difficult to directly search for the decision rule based on \eqref{classification} because of  discontinuity of the indicator functions.
Rich  literature on machine learning methods for multicategory classification problems is now available. In particular, multicategory SVMs  were  studied extensively in recent years (see e.g. Crammer and Singer,\cite{CrammerSinger2001} Lee et al.,\cite{lee2004multicategory} Liu and Yuan,\cite{liu2011reinforced} Zhang and Liu,\cite{zhang2014angle} and Zhang et al.\cite{zhang2016reinforced}).  We adopt  RAMSVM \citep{zhang2016reinforced} under the angle-based classification framework to solve the multicategory weighted classification problem  \eqref{classification} because that RAMSVM is   computationally faster than many other  multicategory SVM algorithms and that it is  Fisher consistent. The R package \texttt{ramsvm}\cite{zhang2016reinforced} is  available to implement RAMSVM using the efficient coordinate descent algorithm.

To this end, we now recast the  multicategory weighted classification problem  \eqref{classification} under the angle-based classification framework.\citep{zhang2014angle, zhang2016reinforced} Consider   $k$ simplex vertices $\bm{V}_{1},\cdots, \bm{V}_{k}$  in $\mathbb{R}^{k-1}$, given by
\begin{equation*}
	\bm{V}_{l}= \begin{cases}(k-1)^{-1 / 2} {\bm{1}}_{k-1}, & \text { for } l=1, \\ -\frac{1+\sqrt{k}}{(k-1)^{3 / 2}} {\bm{1}}_{k-1}+\sqrt{\frac{k}{k-1}} \bm{e}_{l-1}, & \text { for } l=2,\ldots, k,\end{cases}
\end{equation*}
where  ${\bm{1}}_{k-1} \in \mathbb{R}^{k-1}$ is a vector with all elements being 1, and  $\bm{e}_{l} \in \mathbb{R}^{k-1}$ is a zero vector  except that its $l$th element is 1. The $k$ vertices introduce a simplex centered at the origin and each $\bm{V}_l$ has Euclidean norm 1. Moreover, the angle between $\bm{V}_i$ and $\bm{V}_j$, denoted as $\angle\left(\bm{V}_i,\bm{V}_j\right)$, are equal for any $1\leq i\neq j\leq k$. 

Under the angle-based classification framework, the decision rule $\mathcal{D}(\cdot)$ is induced by a $(k-1)$-dimensional vector of classification functions $\bm{f}(\cdot)=\left(f_{1}(\cdot), \ldots, f_{k-1}(\cdot)\right)^\top$ that maps from the covariates space $\mathcal{X}$ to $\mathbb{R}^{k-1}$. To be specific, given any $\bm{X} \in \mathcal{X}$, the decision rule is  determined by  the smallest angle between $\bm{f}(\bm{X})$ and $\bm{V}_{l}$ ($l=1,\ldots,k$): $\mathcal{D}(\bm{X}) = {\arg\min }_{l \in\{1, \ldots, k\}}\angle\left(\bm{f}(\bm{X}), \bm{V}_{l}\right)$, or equivalently, $\mathcal{D}(\bm{X})={\arg\max }_{l \in\{1, \ldots, k\}}\left\langle \bm{f}(\bm{X}), \bm{V}_{l}\right\rangle$,\citep{zhang2014angle}
where  $\langle\cdot, \cdot\rangle$ is the inner product between two vectors.

With this  prediction rule, we consider the following optimization problem  to identify the $k-1$ classification functions:
\begin{equation}\label{decisionF}
	\underset{f\in\mathcal{F}}{\min}\ 
	\sum_{i=1}^n  \underset{\boldsymbol{j}  \in \mathcal{M}_i}{\sum} W^{(\boldsymbol{j})}_i 
	{l}\left(\langle \bm{f}(\bm{X}_i), \bm{V}_{Y_i^{(\boldsymbol{j})}}\rangle\right)+\frac{\lambda}{2}J(f),
\end{equation}
where $\mathcal{F}$ is the function space, $l(\cdot)$ is any  loss function used in the binary SVM algorithms,  $J(\cdot)$ is the penalty term that controls the complexity of the function space, and $\lambda$ is a non-negative tuning parameter to avoid overfitting. The parameter $\lambda$ can be selected by cross-validation.

We employ the reinforced loss function proposed in Zhang et al.\cite{zhang2016reinforced}:
$$
\begin{aligned}
	{l}\left(\langle \bm{f}(\bm{X}_i), \bm{V}_{Y_i^{(\boldsymbol{j})}}\rangle\right)= &
	 (1-\gamma) \sum_{l \neq {Y}^{(\boldsymbol{j})}_i}\left[1+\left\langle \bm{f} \left(\bm{X}_{i}\right), \bm{V}_{l}\right\rangle\right]_{+} \\ + &\gamma\left[(k-1)-\langle \bm{f} \left(\bm{X}_{i}\right), \bm{V}_{{Y}^{(\boldsymbol{j})}_i}\rangle\right]_{+},
\end{aligned}$$
where $\gamma\in[0,1]$ is a pre-specified linear combination parameter. The default choice is $\gamma=0.5$. \textcolor{black}{The motivation of this loss function originates from the argmax rule of angle-based classification problem, where $\langle \bm{f} \left(\bm{X}_{i}\right), \bm{V}_{{Y}^{(\boldsymbol{j})}_i}\rangle$ needs to be the  maximum value among the $k$ inner products $\left\langle \bm{f} \left(\bm{X}_{i}\right), \bm{V}_{l}\right\rangle, l=1,\ldots,k$. The first part is similar to the loss function used in Lee et al.,\cite{lee2004multicategory}  which encourages $\left\langle \bm{f} \left(\bm{X}_{i}\right), \bm{V}_{l}\right\rangle$ to be small for $l\neq {Y}^{(\boldsymbol{j})}_i$. This part indirectly forces $\langle \bm{f} \left(\bm{X}_{i}\right), \bm{V}_{{Y}^{(\boldsymbol{j})}_i}\rangle$ to be large since $\sum_{l=1}^k \left\langle \bm{f} \left(\bm{X}_{i}\right), \bm{V}_{l}\right\rangle=0$. The second part is similar to the hinge loss in the binary SVM literature, which encourages  $\langle \bm{f} \left(\bm{X}_{i}\right), \bm{V}_{{Y}^{(\boldsymbol{j})}_i}\rangle$ to be large directly.} Combining these two loss functions linearly in  problem \eqref{decisionF} guarantees that the resulting optimal decision rule is identical to the one in the multicategory classification problem \eqref{classification} asymptotically under suitable conditions.\citep{zhang2016reinforced}

Next, we discuss the solution of \eqref{decisionF} in details by taking linear decision rules as an example. Assume that $f\in \mathcal{F}$ is linear with  $f_q(\bm{X})=\bm{X}^\top\bm{\beta}_{q},\ q=1,\ldots,k-1$. Here, an intercept term is included. We use the squared $L_2$ norm to formulate the penalty term: $J(f)=\sum_{q = 1}^{k-1}  \|\bm{\beta}_q\|_2^2$, where $\|\bm{\beta}\|_2^2=\bm{\beta}^\top \bm{\beta}$.  Then, the minimization problem \eqref{decisionF} is simplified to a convex optimization problem with finite-dimensional parameters  $\bm{\beta}_{q},\ q=1,\ldots,k-1$, and the dual problem of problem \eqref{decisionF} is a quadratic optimization problem with box constraints by the method of Lagrangian multipliers  \citep{zhang2016reinforced}:
\begin{align*}
	\underset{\alpha_{(i,\boldsymbol{j},l)}}{\min}\ &\frac{1}{2\lambda}\sum_{q=1}^{k-1}
	\left\|\sum_{i=1}^n  \underset{\boldsymbol{j}  \in \mathcal{M}_i}{\sum}
	\alpha_{(i,\boldsymbol{j},{Y}^{(\boldsymbol{j})}_i)}V_{({Y}^{(\boldsymbol{j})}_i,q)}\bm{X}_i
	-\sum_{i=1}^n \underset{\boldsymbol{j}  \in \mathcal{M}_i}{\sum}\sum_{l \neq {Y}^{(\boldsymbol{j})}_i}\alpha_{(i,\boldsymbol{j},l)}V_{(l,q)}\bm{X}_i
	\right\|_2^2\\
	&-\sum_{i=1}^n  \underset{\boldsymbol{j}  \in \mathcal{M}_i}{\sum}\alpha_{(i,\boldsymbol{j},{Y}^{(\boldsymbol{j})}_i)}(k-1)-\sum_{i=1}^n \underset{\boldsymbol{j}  \in \mathcal{M}_i}{\sum}\sum_{l \neq {Y}^{(\boldsymbol{j})}_i}\alpha_{(i,\boldsymbol{j},l)},\\
	\text{s.t.} &\ 0\leq  \alpha_{(i,\boldsymbol{j},l)}\leq A_{(i,\boldsymbol{j},l)}\ (i=1,\cdots, n; \ \boldsymbol{j} \in \mathcal{M}_i; \ l=1,\cdots, k),
\end{align*}
where $V_{(l,q)}$ is the $q$th component of $\bm{V}_l$, and
$A_{(i,\boldsymbol{j},l)}=W^{(\boldsymbol{j})}_i[\gamma I(l=Y_i^{(\boldsymbol{j})})
+(1-\gamma )I(l\neq Y_i^{(\boldsymbol{j})})]$. The dual problem can be solved by the  efficient coordinate descent algorithm. Let $\hat{\alpha}_{(i,\boldsymbol{j},l)} (i=1,\cdots, n;\ \boldsymbol{j} \in \mathcal{M}_i; \ l=1,\cdots, k)$ be the resulting  minimizer of  the dual problem. Then, the estimator of parameter $\bm{\beta}_q$ in the $q$th linear classification function is
$$\hat{\bm{\beta}}_q=\frac{1}{\hat{\lambda}}\left[\sum_{i=1}^n  \underset{\boldsymbol{j}  \in \mathcal{M}_i}{\sum}
\hat{\alpha}_{(i,\boldsymbol{j},{Y}^{(\boldsymbol{j})}_i)}V_{({Y}^{(\boldsymbol{j})}_i,q)}\bm{X}_i
-\sum_{i=1}^n \underset{\boldsymbol{j}  \in \mathcal{M}_i}{\sum}\sum_{l \neq {Y}^{(\boldsymbol{j})}_i}\hat{\alpha}_{(i,\boldsymbol{j},l)}V_{(l,q)}\bm{X}_i
\right].$$
Here, $\hat{\lambda}$ is the cross-validated value of $\lambda$. We illustrate cross-validation in details in the simulation section.

Let $\hat{\bm{f}}(\bm{X})=(\hat{f}_1(\bm{X}),\cdots, \hat{f}_{k-1}(\bm{X}))^\top=
(\bm{X}^\top \hat{\bm{\beta}}_1,\cdots, \bm{X}^\top \hat{\bm{\beta}}_{k-1})^\top$ be the estimated classification function vector given  covariate $\bm{X}$. The  optimal linear  decision rule  by multicategory M-learning is
\begin{equation}\label{optimalITR}
	\widehat{\mathcal{D}}(\bm{X})=
	\underset{l \in\{1, \ldots, k\}}{\arg\max }\left\langle \hat{\bm{f}}(\bm{X}), \bm{V}_{l}\right\rangle
	=\underset{l \in\{1, \ldots, k\}}{\arg\max }\sum_{q=1}^{k-1} \hat{\bm{\beta}}_{q}^\top \bm{X}  V_{(l,q)}.   
\end{equation}

When  the classification functions are possibly nonlinear, we  use  the kernel method with  the  classification functions restricted to the reproducing kernel Hilbert space. We refer the readers to Zhang et al.\cite{zhang2016reinforced} for more details.

In practice,  when the sample size is large and the covariate dimension is moderate or high, it can be computationally intensive to consider all 
permutations $\boldsymbol{j}$ in $\mathcal{M}_i$.  To alleviate the computation burden, one may consider  nearest neighbour  matching   \citep{wu2020matched}  to construct the matched sets, where   $|\mathcal{M}^{(l)}_i|=1$  for $l=1,\ldots,k$.  We then simply write $\widehat{R}_{i}(w)=\widehat{R}^{(\boldsymbol{j})}_{i}(w)$ since there is only one element $\boldsymbol{j}$.  The  proposed multicategory matching-based value function \eqref{valuefun_multi} reduces to
$V_n(\mathcal{D};g)=  \frac{1}{n} \sum_{i=1}^n 
I\{\mathcal{D}(\bm{X}_i)={\arg\max}_{w \in \{1,\ldots,k\}} \widehat{R}_{i}(w)\}  g(R_{j_1},\cdots, R_{j_k}).$

{\color{black} Throughout this paper, we   use nearest neighbour covariate matching with replacement \citep{frolich2004programme} to construct the matched sets in \eqref{multiple_matchedset}, where  the closeness is measured by  Mahalanobis distance. It is known that covariate matching may suffer from the curse of dimensionality, and propensity score matching may be preferred when the covariate dimension is moderate or high. For comparison purpose, we implement  nearest neighbour  matching based on the generalized propensity scores (GPS),\citep{yang2016propensity} where the GPSs are estimated by multinomial logistics regression}. Vector matching,\citep{lopez2017estimation,scotina2019matching,scotina2020matching}
among other recently developed matching methods for multiple treatments, can be used instead.

\subsection{Right-Censored Survival Data}\label{sec3.4}
The proposed multicategory M-learning in Section \ref{sec3.2} and its estimation procedure in Section \ref{sec3.3} can be generalized to handle right-censored survival data. We consider the following nonparametric imputation technique for right-censored survival data.

Let $\tilde{T}$ denote the true survival time and $\tau$ be the end of the study. Assume that the censoring time $C$ is independent of $\tilde{T}$ given $(\bm{X}, A)$. The right-censored  survival time is $T=\min (\tilde{T}, C,\tau)$ and the censoring indicator is $\Delta=I(\tilde{T} \leq C)$. In the survival setting,  the true survival time  is the clinical reward in \eqref{valuefun_multi}, i.e., $R=\tilde{T}$, if all survival times were observed. 

However, the $\tilde{T}$'s are usually not fully observed due to right-censoring. A simple but tentative solution is to replace the survival times with the  conditional expected survival times ${E}(\tilde{T}| \bm{X}, A)$, estimated by  regression or nonparametric  methods. A more principled solution  is that we  impute the censored survival time  as in Cui et al.\cite{cui2017tree}:
\begin{equation*}\label{impute_survival}
	R = \Delta T+(1-\Delta) \hat{E}(\tilde{T} \mid \bm{X}, A, \tilde{T}>T, T),
\end{equation*}
where $\hat{E}(\tilde{T} \mid \bm{X}, A, \tilde{T}>T, T)$ is some nonparametric estimation of  ${E}(\tilde{T} \mid \bm{X}, A, \tilde{T}>T, T)$. That is, we set the reward $R=\tilde{T}$   if the event occurs, and set  $R=\hat{E}(\tilde{T} \mid \bm{X}, A, \tilde{T}>T, T)$  if this subject is censored.

The nonparametric estimation $\hat{E}(\tilde{T} \mid \bm{X}, A, \tilde{T}>T, T)$ can be obtained as follows. We first estimate the survival function ${S}(t |\bm{X},A)=Pr(\tilde{T}>t|\bm{X},A)$  nonparametrically  using random survival forest,\citep{ishwaran2008random} which is obtained by the R package \texttt{randomForestSRC}. Let $\hat{S}(t |\bm{X},A)$ be the estimate. 
Note that 
\begin{align*}
	{E}(\tilde{T} \mid \bm{X}, A, \tilde{T}>T, T)&=T+{E}(\tilde{T}-T \mid \bm{X}, A, \tilde{T}>T, T)\\
	&=T+\frac{\int_T^\tau S(t|\bm{X},A)dt}{ S(T|\bm{X},A)},
\end{align*}
where ${E}(\tilde{T}-T \mid \bm{X}, A, \tilde{T}>T, T)$ is the  mean residual survival time. We  set $\hat{E}(\tilde{T} \mid \bm{X}, A, \tilde{T}>T, T)=T+\frac{\int_T^\tau \hat{S}(t|\bm{X},A)dt}{\hat{S}(T|\bm{X},A)}$. 

Finally, we obtain the optimal ITR by maximizing the  multicategory value function \eqref{valuefun_multi} as in  Sections \ref{sec3.2} and \ref{sec3.3}. When the classification functions are linear, the optimal ITR is  given in \eqref{optimalITR}. If the  classification functions are possibly nonlinear, we use the kernel techniques to identify the optimal ITR.

{\color{black}
\subsection{Theoretical Property}
In this section, we establish the theoretical property of the proposed method.  The proof is given in the Supplemental Material.

\begin{theorem}\label{thm1}
	Suppose that ${\max}_{i\in \{1,\cdots, n\};w \in \mathcal{A}-\{A_i\}} \delta^{(w)}_i \rightarrow 0$ as $n\rightarrow \infty$. Assume that  the weighting function $g(\cdot)$ is  exchangeable and non-negative. Then given any decision rule $\mathcal{D}(\cdot)$ and   function $g(\cdot)$,  the multicategory matching-based value function  satisfies that
	$$V_{n}(\mathcal{D} ; g)\overset{p}{\longrightarrow} V(\mathcal{D} ; g), \ \text{as}\ n\rightarrow \infty,$$
	where the limit $V(\mathcal{D} ; g)$ is defined in the Supplemental Material.  Let $(\widetilde{X}_l,\widetilde{A}_l,\widetilde{R}_l)^{k-1}_{l=1}$ denote the i.i.d. copies of $(X,A,R)$. Define
	\[\Delta_g(\boldsymbol{r},w,x)=E\left[I\{R \geq \underset{l\in\{1,\cdots, k-1\}}{\max} r_l\} \times g(R,r_1,\ldots,r_{k-1}) \mid {A}=w, {X}=x\right],
	\] where $ w=1,\ldots,k$ and $\boldsymbol{r}=(r_1,\ldots,r_{k-1})\in \mathbb{R}^{k-1}$. Given $X=x$ and $g(\cdot)$, let $\mathcal{D}^*(\cdot)$ be the optimal ITR that maximizes
	$ V(\mathcal{D} ; g)$. Then, it can be expressed as
	\[\mathcal{D}^*(x)=\underset{w \in \{1,\ldots,k\}}{\arg \max} \int \Delta_g(\boldsymbol{r},w,x) \mathrm{d} F(\boldsymbol{r}\mid \boldsymbol{A}=\boldsymbol{a}_{[-w]}, \boldsymbol{X}=\boldsymbol{x}), \]
	where $ \boldsymbol{A}=(\widetilde{A}_1,\ldots,\widetilde{A}_{k-1})$, $\boldsymbol{a}_{[-w]}=(1,\ldots,w-1,w+1,\ldots,k)$, $\boldsymbol{X}=(\widetilde{X}_1,\ldots,\widetilde{X}_{k-1})$, $\boldsymbol{x}=(x,\cdots,x)$, and $F(\boldsymbol{r}\mid \boldsymbol{A}={a}_{[-w]}, \boldsymbol{X}=\boldsymbol{x})$ is the joint distribution of $\boldsymbol{R}=(\widetilde{R}_1,\ldots,\widetilde{R}_{k-1})$ given $\boldsymbol{A}=\boldsymbol{a}_{[-w]}$ and $\boldsymbol{X}=\boldsymbol{x}$.
\end{theorem}

Theorem \ref{thm1} generalizes  Theorem 3.1 in Wu et al.\cite{wu2020matched} to the multicategory setting.  Here we give some further remarks on this theorem.

\begin{remark}
	When we choose $g(\cdot)=1$, then $$\Delta_g(\boldsymbol{r},w,x)= Pr(R \geq \underset{l \in\{1,\cdots,k-1\}}{\max}  r_l \mid A=w, X=x),$$ which is the  probability that a subject's outcome $R$ given $A=w$ and $X=x$ is larger than  the rest $k-1$ outcomes if this subject were assigned to  treatment $A\in \mathcal{A}-\{w\}$. Then, the optimal ITR $\mathcal{D}^*(\cdot)$ recommends the treatment that  most probably produces the largest outcome. Analogous to the remark in Wu et al.,\cite{wu2020matched} the constant weighting function makes  multicategory M-learning robust against outliers of the outcomes.
\end{remark}

\begin{remark}
	Suppose the outcomes are non-negative. When we choose $g_1(\cdot)$ in \eqref{gweight1}  or $g_2(\cdot)$  in \eqref{gweight2} as the weighting function and $\boldsymbol{r}=\boldsymbol{0}_{k-1}$,\citep{wu2020matched} then $\Delta_g(\boldsymbol{r},w,x)=c E[R \mid A=w, X=x]$ with a positive constant $c$,  suggesting that  the proposed multicategory M-learning is Fisher consistent as analogous to multicategory O-learning.\citep{lou2018optimal,huang2019multicategory}   Consistency and convergence rate of multicategory M-learning  under more general conditions are currently under investigation.  \citep{Zhang2020sinica}
\end{remark}
}

\section{Simulation Studies}\label{sec4}
In this section, we conduct simulations to compare the finite sample performance of the proposed method and several existing methods in the presence of continuous outcomes. We set the number of treatment arms  $k=4$, the covariate dimension $p=6$, and the total sample size $n=1000$. Each simulation is repeated \textcolor{black}{400} times. {\color{black} Simulations under various settings of $(n,p,k)$ and the scenario of right-censored survival outcomes are included in the Supplemental Material. }

\subsection{Correctly Specified Generalized Propensity Score Model}\label{continuous}

In this subsection, we consider the setting of continuous outcomes where the generalized propensity score model is correct. Two criteria are used to assess the performance of any ITR $\mathcal{D}(\cdot)$ based on an independent test set of size 20000. The first criterion is the empirical value function:
\begin{equation}{\label{empirical_valuefun}}
	\frac{{E}_{n}[R I\{A=\mathcal{D}(\bm{X})\} / \pi(A,\bm{X})]}{{E}_{n}[I\{A=\mathcal{D}(\bm{X})\} / \pi(A, \bm{X})]},
\end{equation}
where ${E}_{n}$ denotes the empirical average and $\pi(a,\bm{x})$ is the generalized propensity score. We calculate \eqref{empirical_valuefun} using true $\pi(a,\bm{x})$ in the test set. The second criterion is the misclassification rate of any ITR $\mathcal{D}(\cdot)$ compared with the true optimal treatment $A^*(\cdot)$: $${E}_{n}[ I\{ \mathcal{D}(\bm{X})\neq A^*(\bm{X})\} ].$$

We consider three  choices of the weighting function $g(\cdot)$ in the multicategory matching-based value function \eqref{valuefun_multi} for   multicategory M-learning: (i). $g(x)\equiv 1$ (Match-g1); (ii).  $g_1(\cdot)$ defined in \eqref{gweight1} (Match-gw1); (iii).  $g_2(\cdot)$ defined in \eqref{gweight2} (Match-gw2).  {\color{black} In the Supplemental Material, we  explore an alternative weighting function based on the difference between the largest outcome and the second largest outcome. Nearest neighbour matching based on covariates and generalized propensity scores using Mahalanobis distance are employed to construct the matched sets in \eqref{multiple_matchedset}, denoted by the suffixes -cov and -gps respectively. Consequently, there are six combinations of  multicategory M-learning. As illustrated in Section \ref{sec3.3}, we implement the proposed multicategory M-learning using the R package \texttt{ramsvm}.  Considering that the classification functions may be nonlinear,  we use Gaussian kernel in the package.  We fix the linear combination parameter $\gamma=0.5$ in the reinforced loss function.  We tune the penalty parameter $\lambda$ in the optimization problem \eqref{decisionF} using three-fold cross-validation, and choose   $\lambda$  that maximizes the average of the empirical value functions \eqref{empirical_valuefun} 
where the unknown generalized propensity score $\pi(a,\bm{x})$ is estimated by multinomial logistics regression.} 

{\color{black} The proposed methods are compared with several competitive approaches: multicategory O-learning (Multi-OL),  \citep{lou2018optimal} multicategory   augmented O-learning (Multi-AOL),\citep{huang2019multicategory} Q-learning, \citep{qian2011performance} and angle-based direct learning (AD-learning). \citep{qi2020multi}  It is noteworthy that Multi-OL, Multi-AOL, and AD-learning require that the  generalized propensity score  $\pi(a,\bm{x})$ is known or estimated. We employ  multinomial logistic regression to  estimate $\pi(a,\bm{x})$.  Similar to the proposed methods, Multi-OL is implemented using the RAMSVM algorithm. We implement  Multi-AOL using the publicly available code by Huang et al,\cite{huang2019multicategory} where we use linear regression for the outcome model. For Q-learning, we consider linear regression on $(1,\bm{X},A,\bm{X}A)$ with $l_1$ penalty, where dummy variables are introduced to encode $A$. AD-learning, which involves multivariate responses linear regression, uses the mixed $l_1/l_2$ norm for the coefficient matrix. Both Q-learning and AD-learning are implemented using the R package \texttt{glmnet}.}

We proceed to describe the data generation process. The six-dimensional covariates $\bm{X}$ are independently generated from $U(0,1)$. Analogous to the setting in Yang et al.,\cite{yang2016propensity}  the four-level treatment $A$ conditioning on $\bm{X}$ is simulated from a multinomial distribution taking values in $\{1,2,3,4\}$ with  $Pr(A = w \mid \bm{X})=\frac{\exp (\bm{X}^\top\bm{\beta}_{w})}{\sum_{w^{\prime}=1}^{4} \exp (\bm{X}^\top \bm{\beta}_{w^{\prime}})}, \  w=1,2,3,4$.
Here, we set $\bm{\beta}_{1}^\top=(1,2,1,1,1,1), \bm{\beta}_{2}^\top=(1,1,2,1,1,1), $ $\bm{\beta}_{3}^\top=(1,1,1,4,1,1) $  and  $\bm{\beta}_{4}^\top=(1,1,-1,1,1,5)$. The multinomial logistic regression model is correctly specified in this situation.

\begin{figure}[t]
	\centering
	\includegraphics[width=1\linewidth]{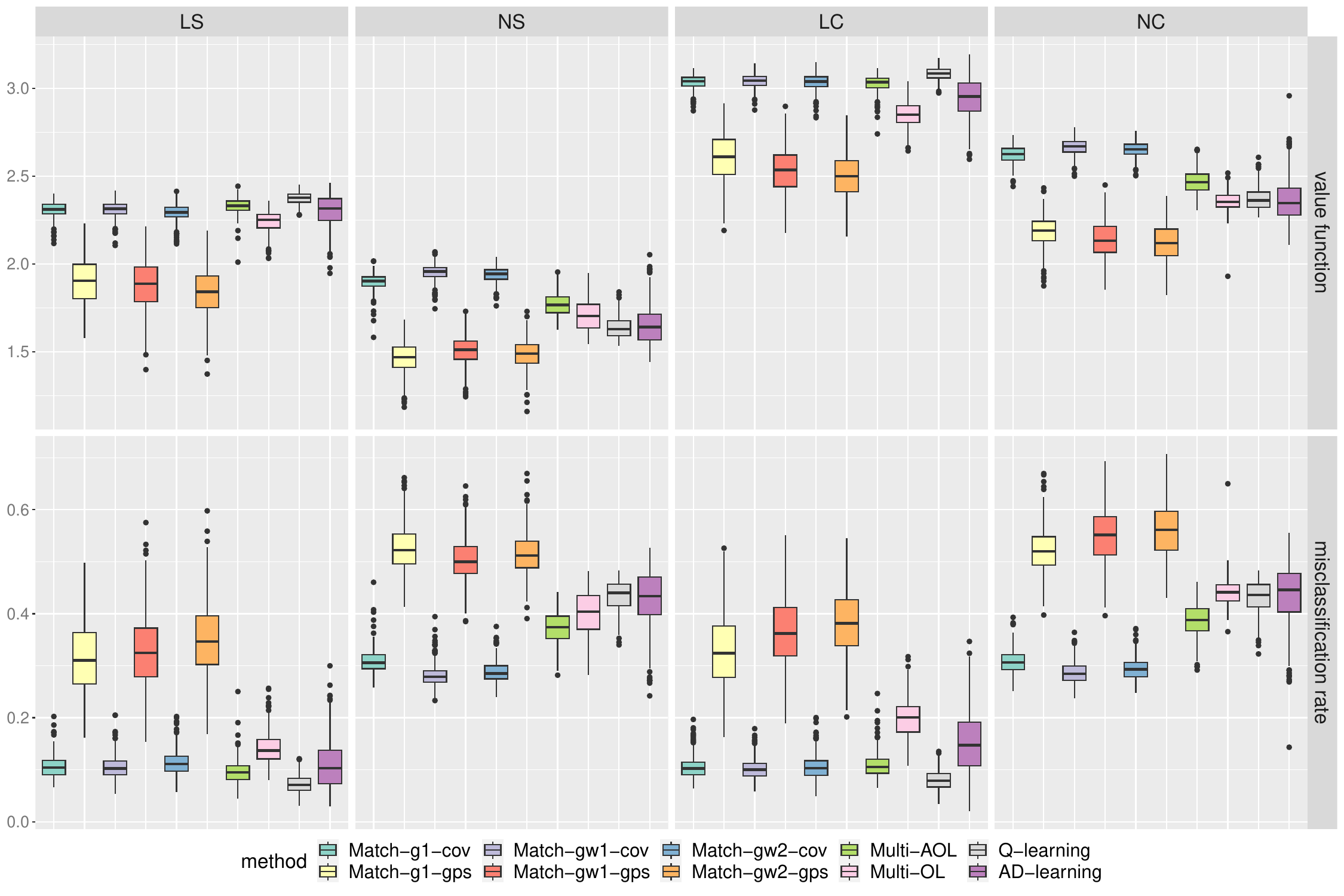}
	\caption{Boxplots for empirical value function and misclassification rate in the presence of continuous outcomes.  The generalized propensity score model is correctly specified.}
	\label{fig:continouscorrectps}
\end{figure}

Next, we describe  the outcome designs as analogous to Lou et al.\cite{lou2018optimal}
Firstly, we consider two scenarios for the true optimal treatment $A^*$ corresponding to linear and nonlinear decision boundaries, respectively. In the scenario of linear decision boundaries, the true optimal treatment $A^*$ is 1 if $X_1>0.5 \text{ and } X_2>0.5$, 2 if $X_1<=0.5 \text{ and } X_2>0.5$, 3 if $X_1<=0.5 \text{ and } X_2<=0.5$, and 4 otherwise. In the scenario of nonlinear decision boundaries, the true optimal treatment $A^*$ is 1 if $0.5\left(X_2-0.5\right)^{2}-X_1+0.7<0,3$
if $0.3<0.5\left(X_2-0.5\right)^{2}+X_1 \leq 0.55,4$ if $0.5\left(X_2-0.5\right)^{2}+$
$X_1 \leq 0.3$, and 2 otherwise.  Finally, the  outcome $R$ is simulated under two scenarios of the main treatment effect. In the scenario of  simple main effect, the outcome $R$ given $A$ and $\bm{X}$ is generated by $R=2 I(A^{*}=A)+X_2$. In the scenario of  complex main effect, the outcome $R$ is generated by
$R=2 I(A^{*}=A)+X_1^{2}+\exp (-X_3-X_4)+\epsilon$, where $\epsilon\sim U(0,1)$.
We consider four scenarios of outcome designs  using different combinations of decision boundary and main effect: (i). linear boundary and simple main effect (LS); (ii). nonlinear boundary and simple main effect (NS); (iii). linear boundary and complex main effect (LC); (iv). nonlinear boundary and complex main effect (NC). {\color{black} The six variants of the proposed methods are compared with the four alternative approaches in each scenario. } Figure \ref{fig:continouscorrectps} summarizes the simulation results.

{\color{black} Figure \ref{fig:continouscorrectps} shows that in the LS and LC scenarios where the decision boundary is linear, the proposed methods based on covariate matching,  Multi-AOL, Q-learning, and AD-learning exhibit comparable performances, with Q-learning  yielding slightly better results. The proposed methods based on GPS matching yield  worse performance, indicating that these methods are not suitable in these scenarios (see the Supplemental Material for the discussion of the situations where  these methods are preferable).  Multi-OL performs poorly, which may be due to small or extreme weights in multiple treatment scenarios. In the NS and NC scenarios where the decision boundary is nonlinear, the proposed methods with covariate matching significantly outperform all other methods. These results suggest that Multi-AOL, Q-learning, and AD-learning rely on the linearity assumption of the decision boundary, whereas the proposed methods with covariate matching are better suited for nonlinear treatment rules. Overall,  multicategory M-learning with covariate matching is preferred, and the variant with the weighting function $g_1(\cdot)$ slightly outperforms the alternatives.}

\subsection{Misspecified Generalized Propensity Score Model}\label{incorrect}

{\color{black} Most of the aforementioned methods  need to specify a model for the  generalized propensity scores before searching for the optimal ITRs, where multinomial logistic regression is commonly employed. Notably,  although multicategory M-learning with covariate matching  does not require propensity score modeling in  the matching-based value function, it tunes the parameter $\lambda$  using the empirical value function \eqref{empirical_valuefun} where $\pi(a,\bm{x})$ is estimated by  multinomial logistic regression. Multi-OL  uses a similar parameter tuning  procedure.} Therefore, model misspecification may deteriorate the performance of  all  methods in comparison.  Since model misspecification  is ubiquitous  in practice, it is worthwhile to evaluate the performance of the methods  when   the  generalized propensity score model is misspecified. 

\begin{figure}[t]
	\centering
	\includegraphics[width=1\linewidth]{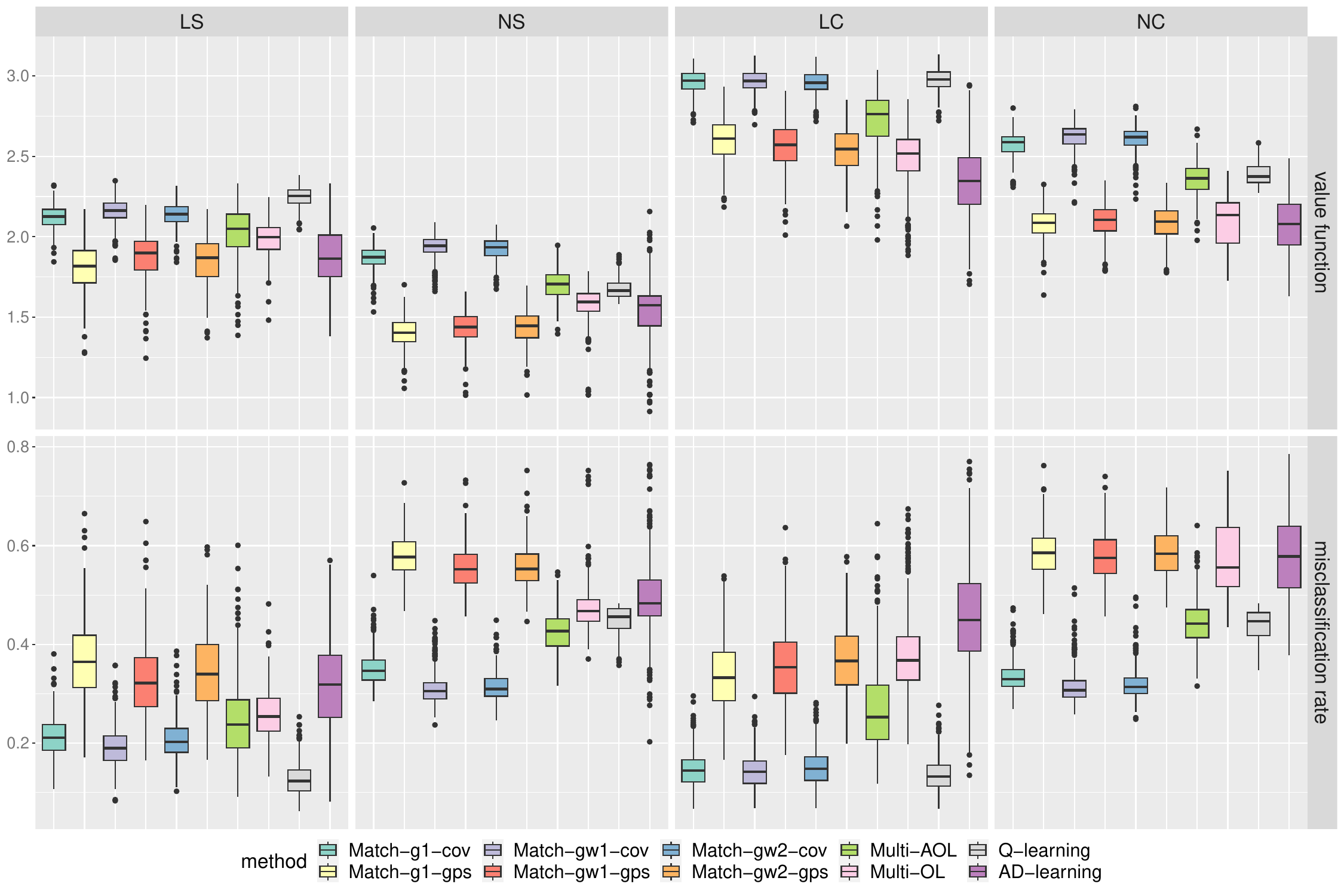}
	\caption{Boxplots for empirical value function and misclassification rate in the presence of continuous outcomes. The generalized propensity score model is misspecified.}
	\label{fig:continousincorrectps}
\end{figure}

Here, we adapt  a simulation setting in Yang et al.,\cite{yang2016propensity} where the four-level treatment $A$ conditioning on $\bm{X}$ is simulated from a multinomial distribution taking values in $\{1,2,3,4\}$ with  
 $Pr (A=w \mid \bm{X})=\frac{\exp ( (\bm{X}^\top \bm{\beta}_{w})^2)}{\sum_{w^{\prime}=1}^{4} \exp (  (\bm{X}^\top \bm{\beta}_{w^{\prime}})^2 )}, \ w=1,2,3,4$.
Here, $\bm{\beta}_{1}^\top=(1,2,1,1,1,1), \bm{\beta}_{2}^\top=(1,1,2,1,1,1), \bm{\beta}_{3}^\top=(1,1,1,2,1,1)$  and  $\bm{\beta}_{4}^\top=(1,1,1,1,1,2)$. Hence  multinomial logistic regression is a misspecified generalized propensity score model. We summarize the simulation outputs in Figure \ref{fig:continousincorrectps}.

{\color{black} Figure \ref{fig:continousincorrectps} reveals that  the proposed methods with covariate matching lead to the best performance in general. The performances become more differentiable in the scenarios with nonlinear decision boundary. Among the proposed methods, Match-gw1-cov achieves the best results, while the variants based on generalized propensity scores matching behave the worst. Although Q-learning slightly outperforms Match-gw1-cov in the LS and LC scenarios, its performance significantly deteriorates in the remaining scenarios. Multi-AOL, Multi-OL, and AD-learning result in unsatisfactory performances due to  misspecification of the generalized propensity score model in the objective functions. It is noteworthy that multicategory M-learning with covariate matching is more robust against model misspecification when tuning $\lambda$ using the empirical value function \eqref{empirical_valuefun}. This finding is consistent with the conclusion of the simulation studies in Wu et al.\cite{wu2020matched}}

\section{Analysis of a Hepatocellular Carcinoma Study}\label{sec5}
Hepatocellular carcinoma, the most common type of liver cancer, accounts for about 75\% incidences of all liver cancers. A  multi-arm multi-center observational study of  HCC enrolled 537  patients within the Milan criteria  between  2009 and  2019  in Tianjin, China,\citep{zhang2020EASL,zhang2020gut} where three types of treatments LT, LR and LA were applied as described in the introduction. Among them, 172, 191 and 174 patients were assigned to   LT, LR and LA, respectively.

Due to heterogeneity and complexity of HCC and a lack of evidence-based studies,   controversies  exist on the efficacy of the three treatments  for the HCC disease 
within the Milan criteria.\citep{zhang2020gut} One goal of this HCC study was to provide a prognosis prediction method for individualized tailored treatment allocation and offer guidance to clinicians during multidisciplinary decision-making. In this analysis, we  estimate the optimal individualized treatment using the aforementioned methods. {\color{black} Although these methods were evaluated in simulation studies with continuous outcomes, they can be adapted for survival data. The implementation details are provided in the Supplemental Material. In addition, we include the Cox model with covariates $\tilde{\bm{X}}=(\bm{X},A,\bm{X}A)$  as a competing method, picking the best treatment with minimal risk score  $\exp{(\tilde{\bm{X}}^{\top}{\bm{\beta}}_{cox})}$.\citep{therneau2000cox}  The  coefficients  ${\bm{\beta}}_{cox}$ are estimated by the R package \texttt{survival}.}

The outcome of interest is the survival time after receiving the treatment. The median follow-up time is 57.2 months. We set $\tau=2000$ days. The censoring rate is  73.7\%. The baseline covariates include age, gender, body mass index, albumin-bilirubin score, aminotransferase-to-platelet ratio index, tumor group, portal hypertension, type 2 diabetes, MELD score, child-pugh grade, and alpha-fetoprotein grade. We standardize continuous covariates and introduce dummy variables for categorical covariates, and the covariate dimension is $p=14$ in our analysis. {\color{black} It is worth noting that the data set originates from a multi-center observational study where each center has its own specialties in treating HCC patients. The treatment allocation is closely related to baseline clinical characteristics, and heterogeneity in study specifications across centers is likely to present. Analogous to the assumption made by Athey and Wager,\cite{athey2021policy} we adopt the selection-on-observables strategy and  assume that controlling  the baseline covariates is sufficient to adjust for the different study specifications across centers. }

\begin{figure}[t]
	\centering
	\includegraphics[width=1\linewidth]{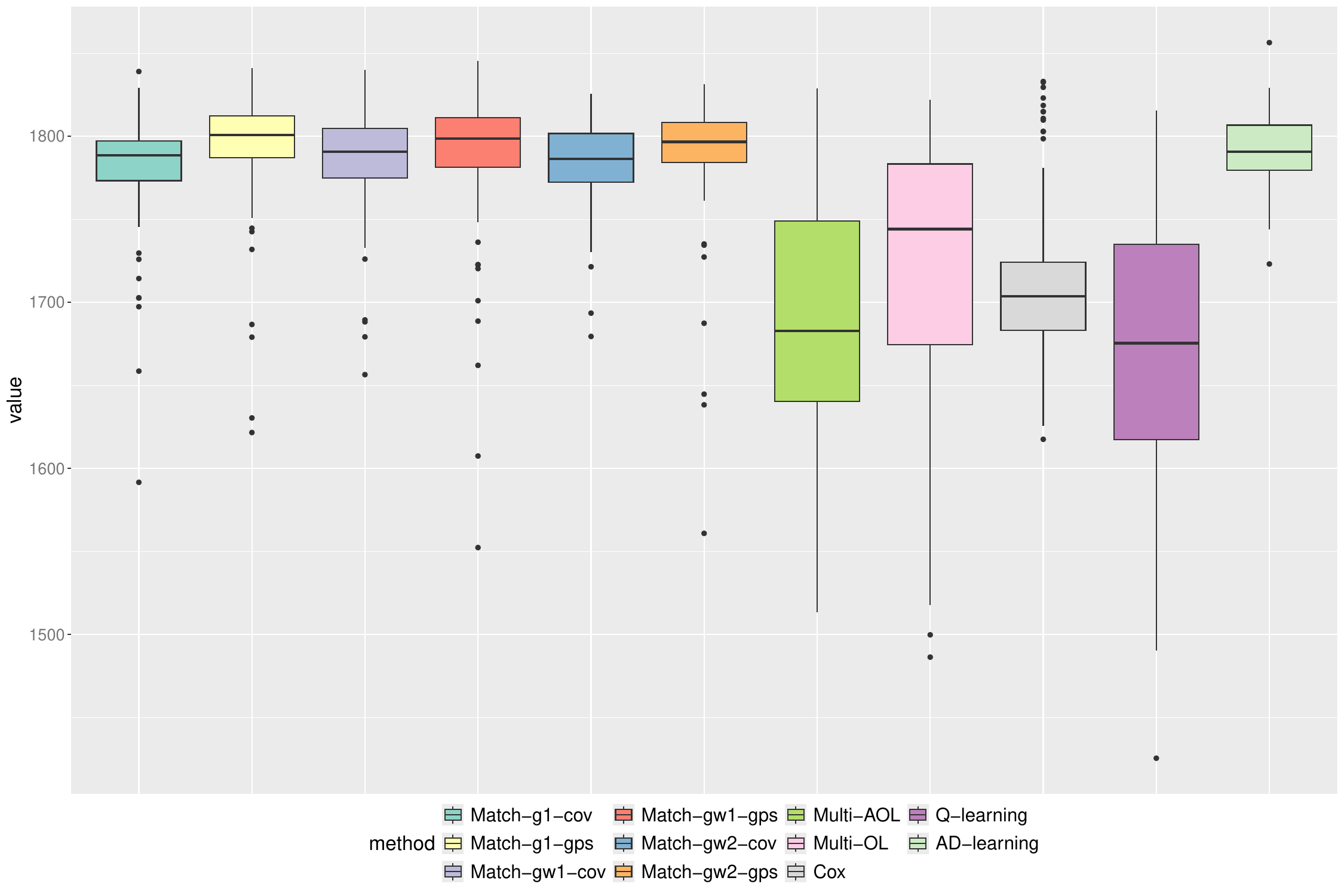}
	\caption{Boxplot for cross-validated empirical values of the expected survival time.}
	\label{fig:valuefuncovgpsmatch}
\end{figure}

We calculate the empirical value function \eqref{empirical_valuefun} for each ITR obtained by these methods, where  the outcome $R$ is obtained by nonparametric imputation using random survival forest, and
the generalized propensity score is estimated by multinomial logistics regression.  Higher value of the empirical  value function means  longer expected survival time if the treatment rule were implemented. Analogous to the cross-validated analysis in Zhao et al.,\cite{zhao2015doubly} we randomly divide the whole data set into five parts with equal sample sizes. Among them, four parts are used as training data to estimate the optimal treatment rules and the remaining fifth part is used as testing data to calculate the empirical value function for each method. The partition of training and testing data is then repeated five times; each time, a different part is treated as testing data. This process results in five empirical  value functions based on the testing data for each ITR, and we record the average value.  We repeat this procedure 100 times and obtain the mean and standard deviation of  the average values. The cross-validated results of the empirical value functions are presented  in Figure \ref{fig:valuefuncovgpsmatch} and  Table \ref{realdata_table}.

{\color{black} The proposed methods generally yield longer expected survival times and smaller standard deviations compared to most of the other methods.  Matching based on generalized propensity scores is generally better than matching based on the covariates, suggesting that generalized propensity scores matching is preferable when the covariate dimension is moderate or high compared to the sample size.\citep{yang2016propensity}  The empirical values with different weighting functions are comparable, and Match-g1-gps  achieves the best performance. This result suggests that the constant weighting function is a suitable choice for this data set, providing robustness against outliers in the outcomes for the proposed method. }

{\color{black} The value functions of Multi-AOL and Multi-OL are smaller than those of the proposed methods. Meanwhile, the standard deviations are significantly larger. A possible explanation is the presence of extreme weights brought by the estimated generalized propensity scores. Additionally, Multi-AOL performs worse than Multi-OL due to that the true decision boundary is possibly highly nonlinear whereas the public code for Multi-AOL \citep{huang2019multicategory} is restricted to the linear case. In contrast, we implement  Multi-OL using Gaussian kernels under the RAMSVM framework. The Cox model does not work well in this data set possibly because the true underlying distribution violates the proportional hazards assumption and the functional form is misspecified. It is noteworthy that AD-learning achieves the second-best value function with the smallest standard deviations, partly due to its equivalence to fitting a weighted Cox model with modified covariates,\citep{qi2020multi} which could be appropriate for this data set.}

Finally, to quantify the improvement using individualized treatment rules, we consider   non-individualized treatment rules: the original treatment assignments in the data set,  the  universal rules that assign all patients to the same treatment arm, and the random rule that arbitrarily assigns patients to one of the three treatment arms.\citep{chen2018estimating}  {\color{black} The value function of the original treatment assignments is 1521.7. When the universal rule with  LT, LR, or LA is applied to all patients,   the value function is 1815.1, 1333.3, or 1495.4, respectively. We also calculate the value function under the random rule repeatedly, and the mean and standard deviation are respectively 1498.6 and 113.1. } The value functions of the proposed methods are similar to the universal rule with LT, indicating that most of the patients are recommended to LT by the proposed methods and that LT is the optimal treatment for most patients, which is consistent with the findings in this HCC study.\citep{zhang2020EASL,zhang2020gut}  {\color{black} As indicated by the simulation results in Section S2 of Supplemental Material, the proposed method based on generalized propensity scores matching is preferable in the scenario with a dominant class.}

\begin{table}[t]
\color{black}
	\centering
	\caption{Cross-validated  empirical values of the expected  survival time. The  standard deviation is inside the bracket. COV and GPS represent the proposed method based on covariate matching and generalized propensity scores matching, respectively. }
	\label{realdata_table}
	\resizebox{\textwidth}{!}{
		\begin{tabular}{lcccccccc}
			\toprule
			Matching Method & Match-g1   & Match-gw1 & Match-gw2 & Multi-AOL & Multi-OL & Cox & Q-learning & AD-learning\\
			\midrule
			COV & 1780.8 (34.5) & 1785.9 (30.8) & 1783.6 (26.8) &\multirow{2}{*}{1683.3 (75.1) }&\multirow{2}{*}{1718.1 (79.1)} &\multirow{2}{*}{1711.3 (49.9)} &\multirow{2}{*}{1674.7 (73.5)} &\multirow{2}{*}{1792.0 (\textbf{20.8})}\\
			GPS & \textbf{1793.2} (35.2) & 1789.0 (42.3) & 1788.6 (38.5) & & &  \\
			\bottomrule
	\end{tabular}}
\end{table}

\section{Discussion}\label{sec6}
We proposed multicategory M-learning, a machine learning method based on matching to estimate optimal ITRs in observational studies with multiple treatments. The optimization problem was solved using the RAMSVM algorithm. With nonparametric imputation of the censored observations using random survival forest, the proposed method was  extended to handle survival outcomes.  We showed theoretical property for the proposed method. Several simulation studies were conducted to evaluate the finite performance of our method compared to existing competitive approaches. Finally, we applied these methods to  analyze the HCC data and further confirmed that the proposed method is promising. 

{\color{black}
The matched set is flexible for leveraging different techniques to improve its quality. For the distance metric, one may focus on  balancing some key covariates if domain knowledge is given.\citep{wu2020matched} Mahalanobis matching on these covariates within propensity score calipers can  be considered to improve the matches. \citep{stuart2010matching} For categorical covariates, introducing dummy variables is necessary, although a large number of dummy variables may be needed for covariates with many categories. To handle the relatively high-dimensional covariates, distance metrics based on dimension reduction techniques including propensity score and prognostic score may be desirable. Antonelli et al.\cite{antonelli2018doubly} combined both scores to construct double robust matching estimation. It is meaningful to consider generalized prognostic score in multicategory M-learning. 

Although we used nearest neighbour matching to develop the matched set throughout this paper,  the number of neighbours can be adjusted, and a data-driven approach like cross-validation can determine the optimal choice. However, using a larger number of neighbours significantly increases the computational burden.  Various matching techniques can be used instead, including vector matching, optimal matching, and full matching.

The multicategory matching-based value function may be improved or extended in several ways. One limitation is that only the nearest neighbour has been used for matching in the implementation. Wu et al.\cite{wu2020using} used all possible matched pairs, weighting each pair's contribution by a kernel based on covariates similarity. Modifying the value function to a kernel-weighted version may potentially enhance performance.  The weighting function $g(\cdot)$ provides a flexible choice to extract information from outcome differences. Although $g(\cdot) \equiv 1$ is the most robust choice against outcome outliers, extensive simulation studies show no significant performance difference among various weights, though $g_1(\cdot)$ performs slightly better. Other versions of $g(\cdot)$ can be explored, and a data-driven strategy for selecting the optimal weighting function may improve performance. \citep{wu2020matched} The proposed method may deteriorate when the outliers of labels are  present, and it is sensible to robustify the performance by using truncated hinge loss function.\citep{zhang2018robust} For the survival setting, we use nonparametric imputation to handle censored observations. Alternative choices  can be considered. \citep{bai2017optimal,jiang2017estimation,qi2020multi,xue2021multicategory}}

\begin{acks}
The authors would like to thank Ningning Zhang, Zhongyang Shen and Wei Lu of Tianjin Medical University Cancer Institute and Hospital and Tianjin First Center Hospital for a stimulating collaboration in hepatocellular carcinoma research.
\end{acks}

\begin{dci}
The author(s) declared no potential conflicts of interest with respect to the research, authorship and/or publication of this article.
\end{dci}

\begin{funding}
Ying Wu's research is  supported by the National Natural Science Foundation of China (Grant 11701295). Ying Yan's research is  supported by the National Natural Science Foundation of China (Grant 11901599; Grant  71991474).
\end{funding}

\begin{sm}
Supplemental material for this article is available online. The code to implement multicategory M-learning  is available in GitHub  via the following link: \verb"https://github.com/XuqiaoLi/MultiMatchITR".
\end{sm}

\end{document}

\maketitle

\section{Simulation Studies with Survival Outcomes}\label{survival}

In this section, we consider the setting of survival outcomes. We  use the empirical value function  and the misclassification rate with $R=\tilde{T}$ as the criteria to assess the performance of any ITR $\mathcal{D}(\cdot)$ based on a large test set. 

{\color{black} We compare the proposed methods, including six variants, with Multi-AOL, Multi-OL, Q-learning, AD-learning, and Cox regression model. When we tune the parameters based on the  empirical value function, we use nonparametric imputation to obtain $R$.  Note that Multi-OL and Multi-AOL are not intended for right-censored survival data, so we utilize the same imputation strategy \citep{cui2017tree}. For Q-learning, we consider the version adjusted with censoring weights \citep{Goldberg2012qlearning,zhao2015doubly}. We fit the log of the true survival time $\tilde{T}$ on $(1,X,A,XA)$ with $l_1$ penalty using the R package \texttt{glmnet}, where the conditional survival function of $C$ given $(A,X)$ is estimated by Cox model. We use accelerated proximal gradient algorithm to implement AD-learning in survival setting, which is equivalent to fitting a weighted Cox model with modified covariates \citep{qi2020multi}. Finally, we include Cox model with covariates $(X,A,XA)$  as a competing method, which picks the best treatment with minimal risk score.}

\begin{figure}[t]
	\centering
	\includegraphics[width=1\linewidth]{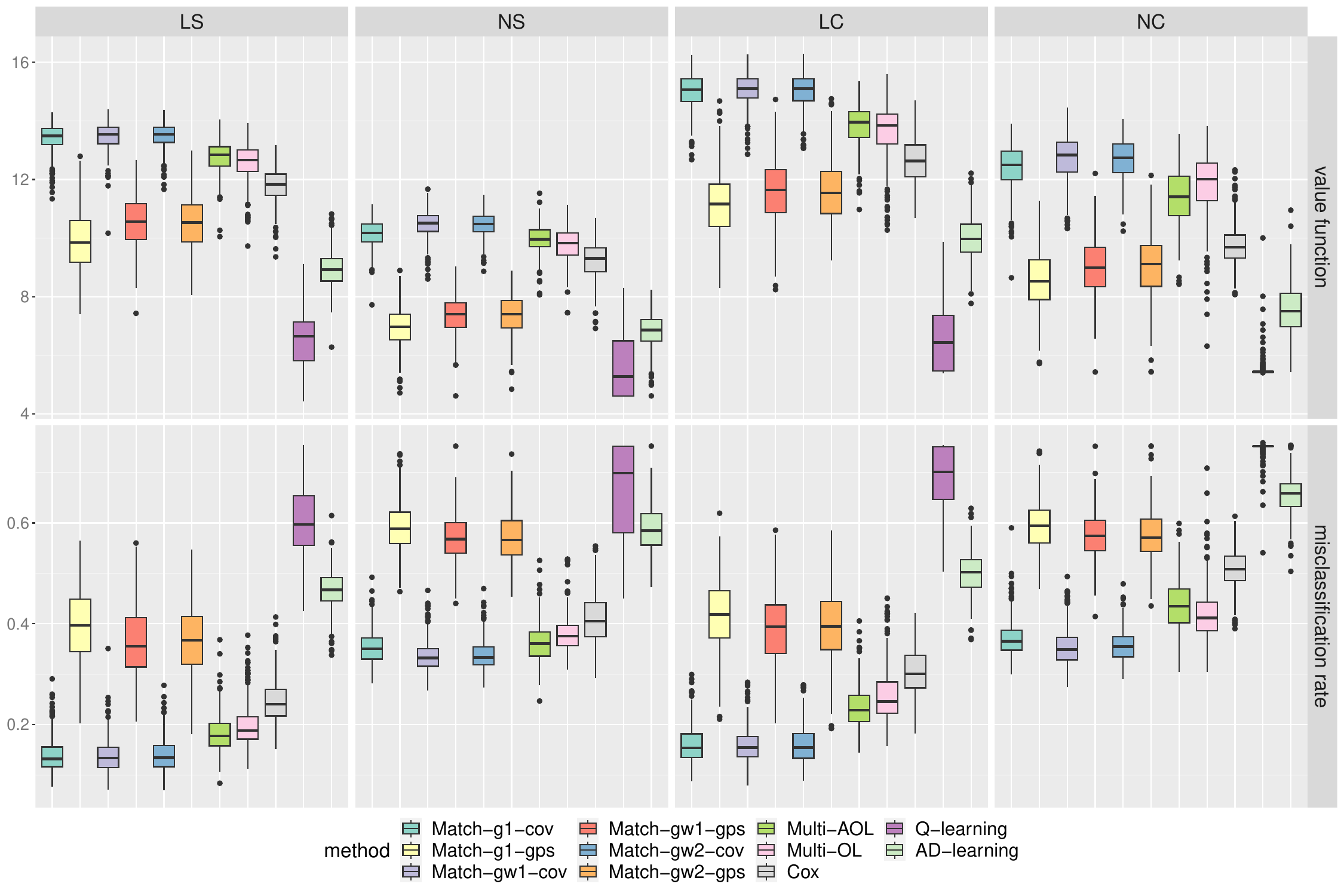}
	\caption{Boxplots for empirical value function and misclassification rate 
		in the presence of survival outcomes, where the true survival time $\widetilde{T}$ is generated from a stratified Cox model. The generalized propensity score model is correctly specified.}
	\label{fig:survivalcoxph2}
\end{figure}

We continue to use the correctly specified propensity score model and   the optimal treatment $A^*$ with linear or nonlinear decision boundary. We consider the simple main effect setting with  
$R_{\tilde{T}}=2 I(A^{*}=A)+X_2$ and  the complex main effect setting with  
$R_{\tilde{T}}=2 I(A^{*}=A)+X_1^{2}+\exp (-X_3-X_4)$, and then 
use a stratified Cox model with treatment-specific  baseline hazard functions to generate the true survival time $\tilde{T}$:
\begin{equation*}
	\lambda(t|A,X)=\lambda_{0}(t|A)\exp(-R_{\tilde{T}}),
\end{equation*}
where the baseline hazard functions in the four treatment arms are  $\lambda_{0}(t|A=1)=1$, $\lambda_{0}(t|A=2)=5I(0<t\le0.3)+2I(0.3<t\le8)+0.7I(t>8)$, ${\lambda _0}(t|A=3) = I(0 < t \le 0.25)\exp ( - 0.3t)+ I(0.25 < t \le 0.75)\exp ( - 0.075)+I(t > 0.75)\exp ( 0.3(t - 1))$, and $\lambda_0(t|A=4) = I(0 < t \le 1)\exp (0.5t) +I(t > 1)\exp ( - 0.5(t - 2)) + 2$, respectively. Note that the proportional hazards assumption is violated. The end of study $\tau$ is fixed to be 12.1, and the censoring time $C$ is generated from an exponential distribution with parameter 0.09 to induce around $31\%$ censoring rate. Figure \ref{fig:survivalcoxph2} presents the simulation outputs.

{\color{black} Figure \ref{fig:survivalcoxph2} shows that the proposed methods based on covariate matching outperform the other methods across all scenarios, with the variants using three different weighting functions behaving similarly. These results suggest that using nonparametric imputation for the proposed method is promising. The performances of the proposed methods based on generalized propensity scores matching, Multi-AOL, and Multi-OL are consistent with the patterns presented in Figure 1. In contrast, Q-learning and AD-learning exhibit poor performance in all scenarios due to the misspecification of the survival time model. It is noteworthy that in the NS scenario, the Cox regression model is comparable with the other methods, despite the violation of the proportional hazards assumption and the misspecification of the functional form. This simulation demonstrates that the Cox regression model is quite robust, which aligns with the findings in \cite{zhao2015doubly,cui2017tree}.}

\begin{figure}[t]
	\centering
	\includegraphics[width=1\linewidth]{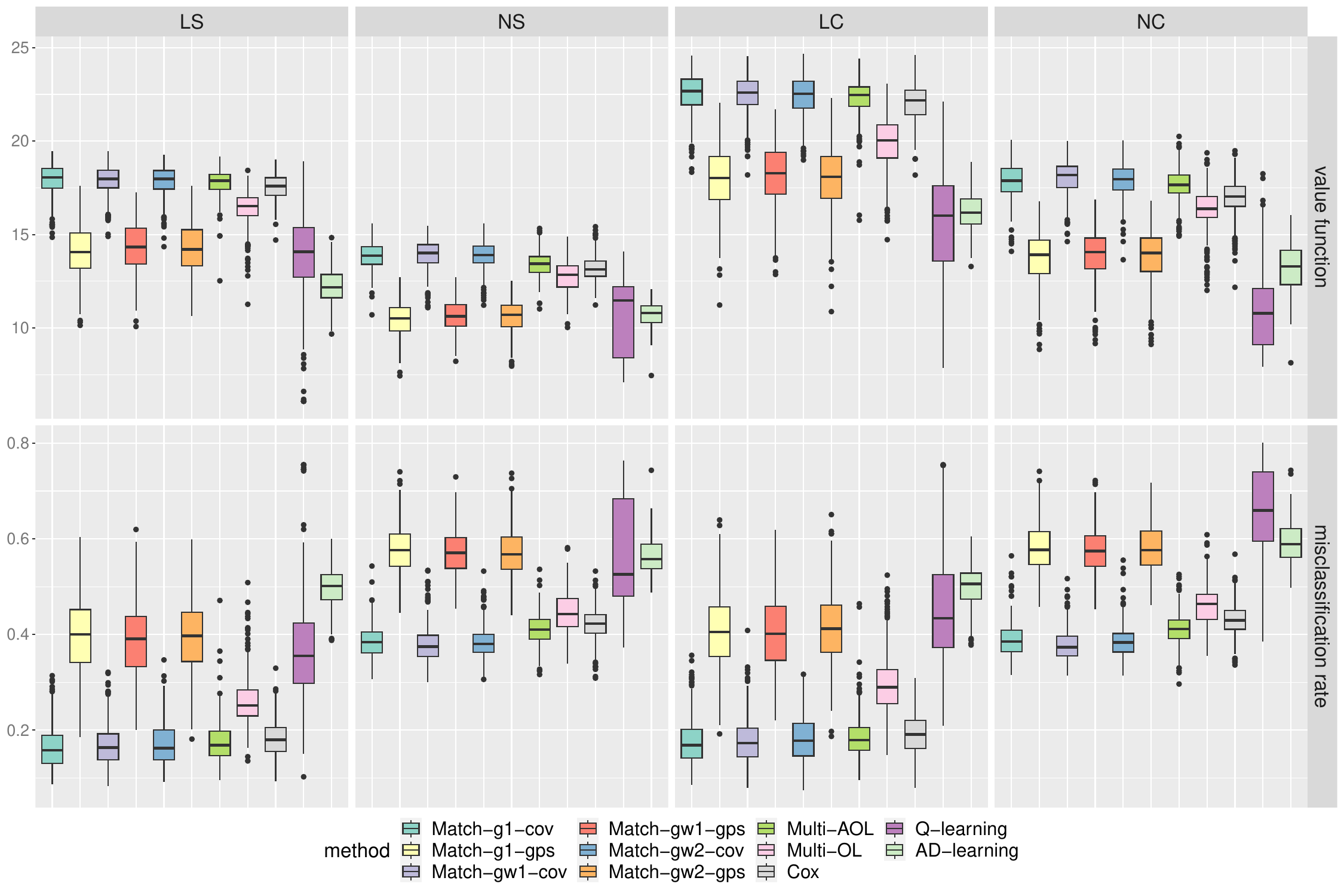}
	\caption{Boxplots for empirical value function and misclassification rate 
		in the presence of survival outcomes, where the true survival time $\widetilde{T}$ is generated from an accelerated failure time model. The generalized propensity score model is correctly specified.}
	\label{fig:survival_lognormal}
\end{figure}

Moreover, we conduct another simulation study where the survival times are generated by an accelerated failure time model given as: 
$$\log(\tilde{T})=R_{\tilde{T}}+\epsilon,$$ where $\epsilon$ is a noisy term generated independently from $N(0,1)$. The end of study $\tau$ is fixed to 11.7, and the censoring time $C$ is generated from an exponential distribution with parameter 0.08 to induce  around $33\%$ censoring rate. The results presented in Figure \ref{fig:survival_lognormal}  are similar to those in  Figure \ref{fig:survivalcoxph2} in general.

\begin{figure}[H]
	\centering
	\includegraphics[width=1\linewidth]{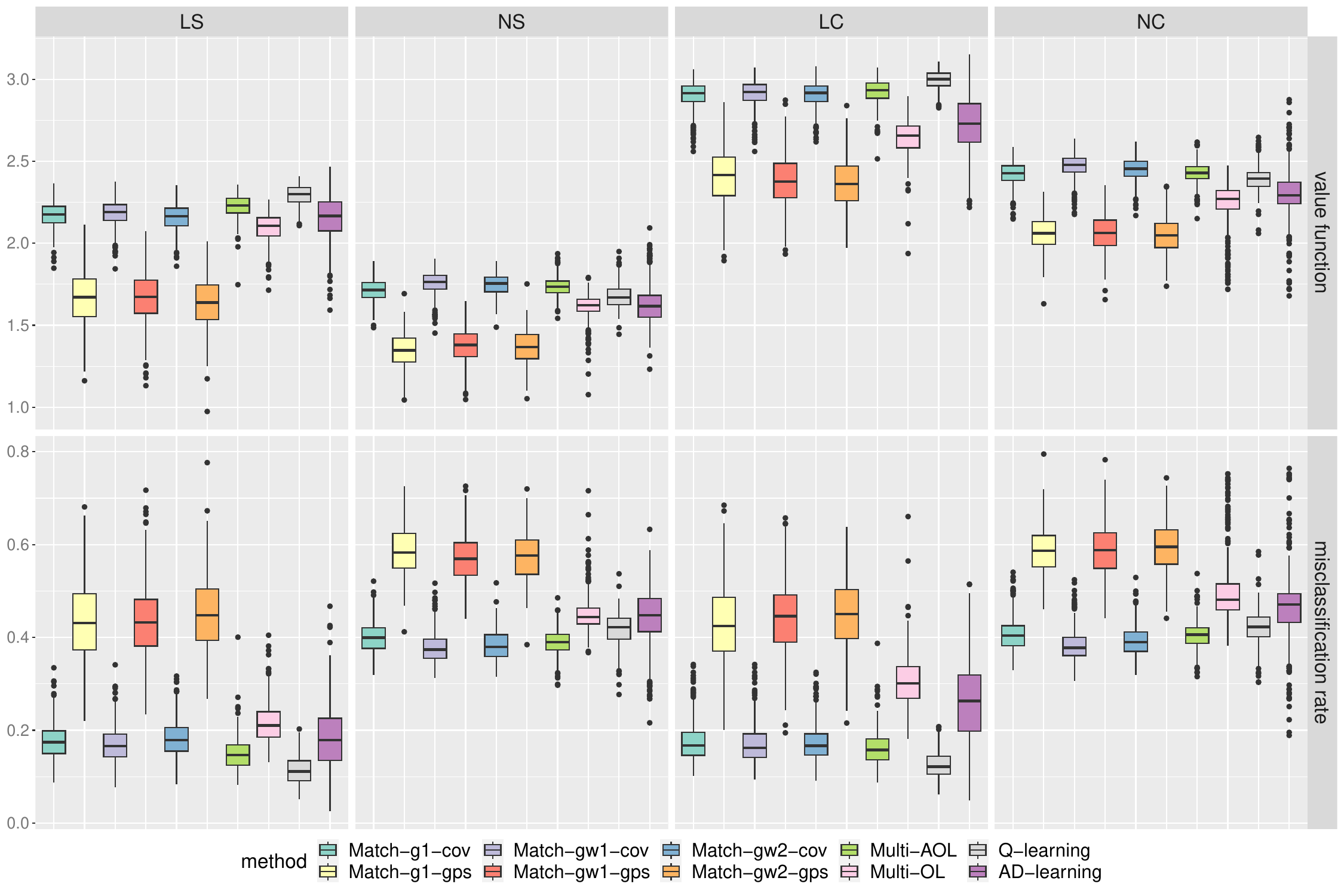}
	\caption{Boxplots for empirical value function and misclassification rate in the presence of continuous outcomes. The simulation setup is the same as in Section 4.1, except the sample size $n$ is reduced to 400.}
	\label{fig:continouscorrectps-n400}
\end{figure}

\begin{figure}[H]
	\centering
	\includegraphics[width=1\linewidth]{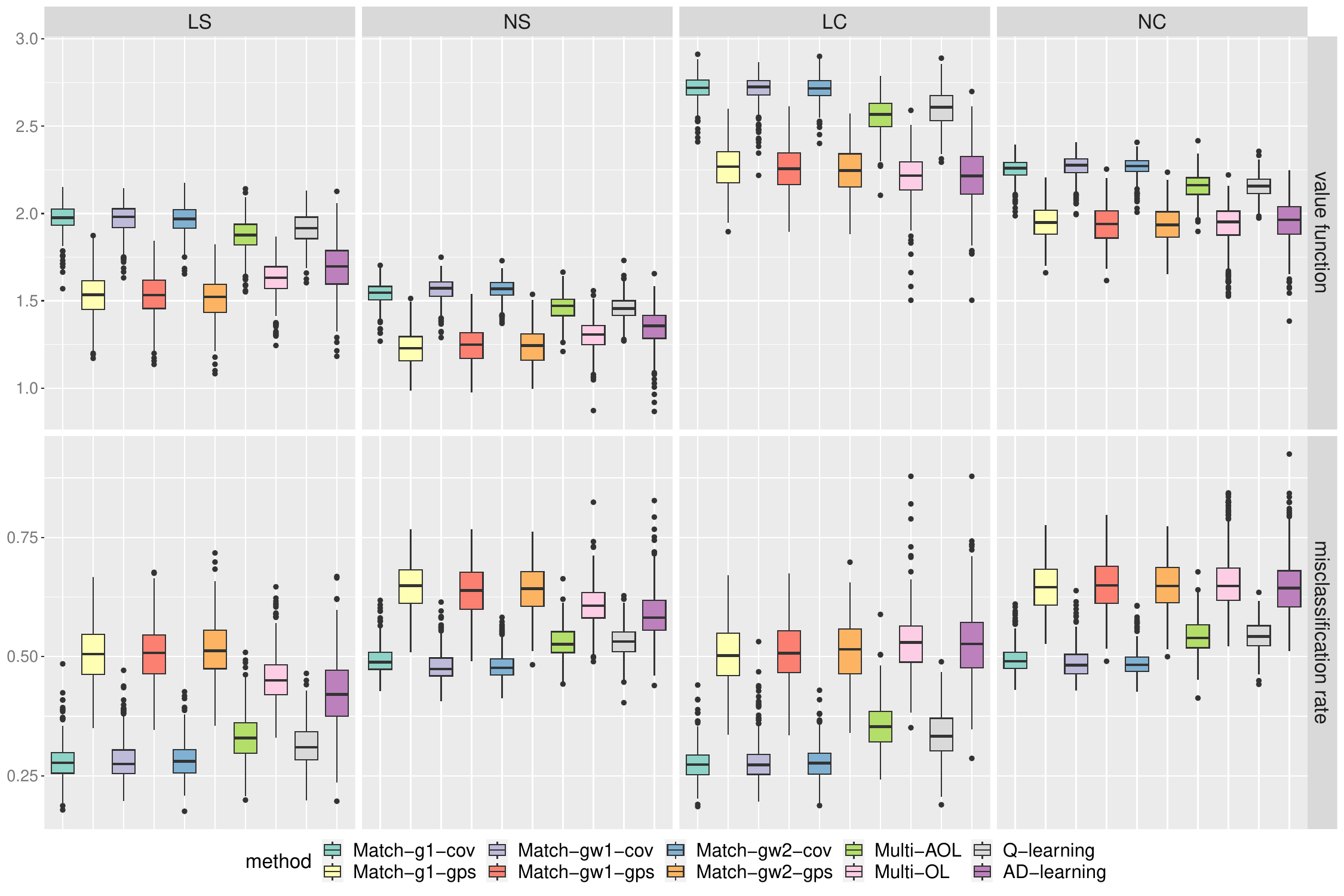}
	\caption{Boxplots for empirical value function and misclassification rate in the presence of continuous outcomes. The simulation setup is adapted from Section 4.1, with the number of treatment arms $k$ increased to 8.}
	\label{fig:continouscorrectps-k8}
\end{figure}

\begin{figure}[H]
	\centering
	\includegraphics[width=1\linewidth]{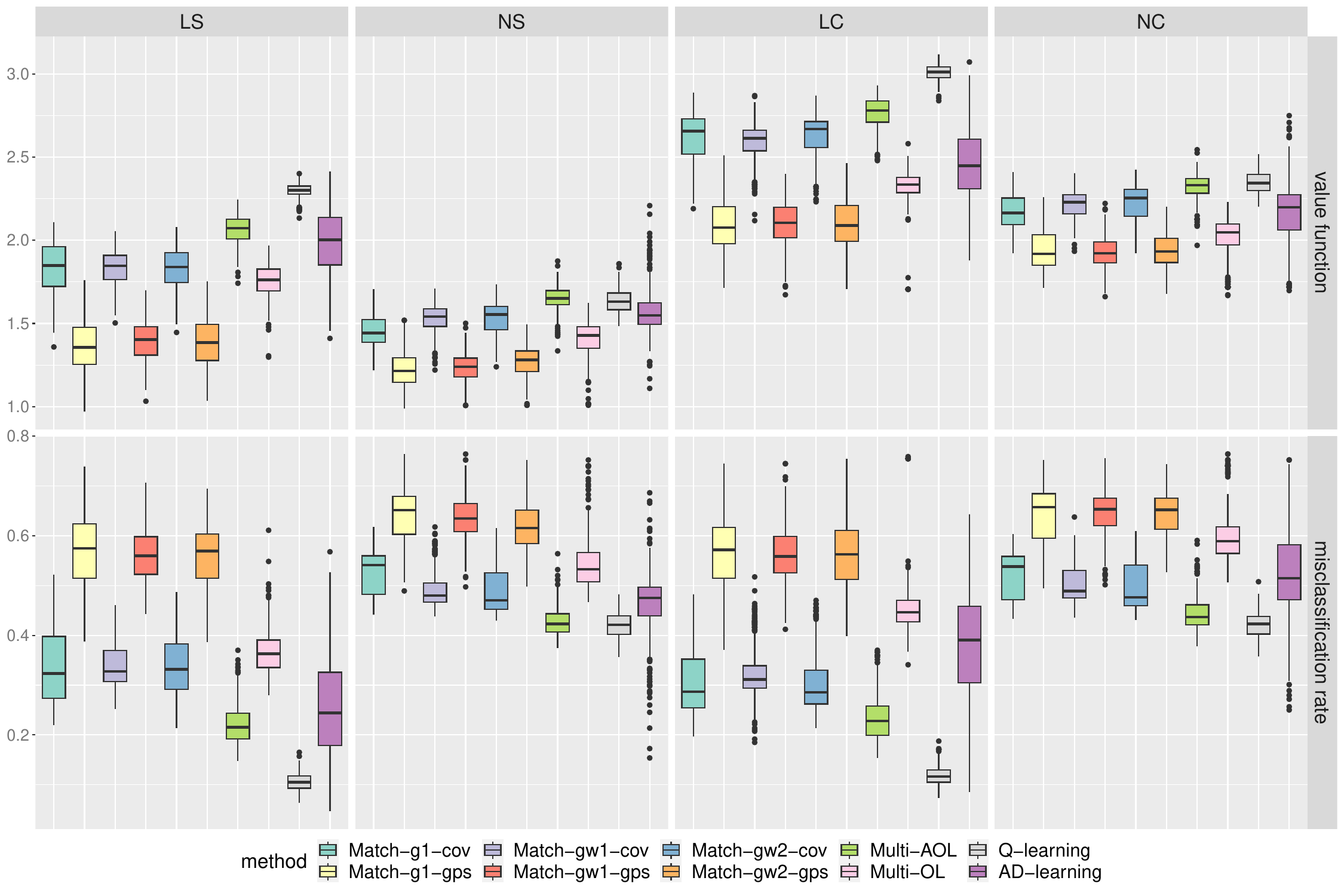}
	\caption{Boxplots for empirical value function and misclassification rate in the presence of continuous outcomes.  The simulation setup is adapted from Section 4.1, with the covariate dimension $p$ increased to 18.}
	\label{fig:continouscorrectps-p18}
\end{figure}

{\color{black}
\section{Simulation Studies with Different Configurations}\label{different-npk}
In Section 4.1, we have investigated the performance of all methods with correctly specified generalized propensity score model, fixing the sample size $n=1000$, the number of covariate dimension $p=6$, and the number of treatment arms $k=4$. In this section, we explore the impacts of varying these factors based on the setup in Section 4.1.

First, we consider the setup in Section 4.1 with a smaller sample size $n=400$. As one reviewer points out, observational studies with small sample sizes are commonly seen. Figure \ref{fig:continouscorrectps-n400} demonstrates that all the methods deteriorate with smaller value functions, higher misclassification rates, and larger variability. However, the results still follow the same pattern as in Figure 1. Q-learning slightly outperforms the proposed methods with covariate matching in the scenarios with linear decision boundary, while our methods demonstrate superior performance in the scenarios with nonlinear decision boundary. 

Second, we consider the setup in Section 4.1 with the number of treatment arms increased to $k=8$. The treatment $A$ conditioning on ${X}$ is simulated from a multinomial distribution taking values in $\{1,\ldots,8\}$ with  $Pr(A = w \mid {X})=\frac{\exp ({X}^\top{\beta}_{w})}{\sum_{w^{\prime}=1}^{4} \exp ({X}^\top {\beta}_{w^{\prime}})}, \  w=1,\ldots,8$, where we set $\beta_{5}=\beta_{1}$, $\beta_{6}=\beta_{2}$, $\beta_{7}=\beta_{3}$, and $\beta_{8}=\beta_{4}$. Here, $\beta_{1},\ldots,\beta_{4}$ follow the same configuration as in Section 4.1. In addition, we also consider the linear and nonlinear decision boundaries for the true optimal treatment $A^*$, which are presented in Table \ref{8treatment-decisionboundary}.

\begin{table}[htbp]
\color{black}
\centering
\caption{The true optimal treatment $A^*$ for linear and nonlinear decision boundaries}
\resizebox{\textwidth}{!}{
\begin{tabular}{c|c|c}
\hline
\textbf{$A^*$} & \textbf{Linear Decision Boundary} & \textbf{Nonlinear Decision Boundary} \\ \hline
1 & $X_1 > 0.5, X_2 > 0.5,$ and $X_3 > 0.5$ & $0.5(X_2 - 0.5)^2 - X_1 + 0.7 < 0$ and $X_3^2 > X_4$ \\ 
2 & $X_1 \leq 0.5, X_2 > 0.5,$ and $X_3 > 0.5$ & $0.5(X_2 - 0.5)^2 - X_1 + 0.7 \geq 0$, $0.5(X_2 - 0.5)^2 + X_1 > 0.55$, and $X_3^2 > X_4$ \\ 
3 & $X_1 \leq 0.5, X_2 \leq 0.5,$ and $X_3 > 0.5$ & $0.3 < 0.5(X_2 - 0.5)^2 + X_1 \leq 0.55$ and $X_3^2 > X_4$ \\ 
4 & $X_1 > 0.5, X_2 \leq 0.5,$ and $X_3 > 0.5$ & $0.5(X_2 - 0.5)^2 + X_1 \leq 0.3$ and $X_3^2 > X_4$ \\ 
5 & $X_1 > 0.5, X_2 > 0.5,$ and $X_3 \leq 0.5$ & $0.5(X_2 - 0.5)^2 - X_1 + 0.7 < 0$ and $X_3^2 \leq X_4$ \\ 
6 & $X_1 \leq 0.5, X_2 > 0.5,$ and $X_3 \leq 0.5$ & $0.5(X_2 - 0.5)^2 - X_1 + 0.7 \geq 0$, $0.5(X_2 - 0.5)^2 + X_1 > 0.55$, and $X_3^2 \leq X_4$ \\ 
7 & $X_1 \leq 0.5, X_2 \leq 0.5,$ and $X_3 \leq 0.5$ & $0.3 < 0.5(X_2 - 0.5)^2 + X_1 \leq 0.55$ and $X_3^2 \leq X_4$ \\ 
8 & $X_1 > 0.5, X_2 \leq 0.5,$ and $X_3 \leq 0.5$ & $0.5(X_2 - 0.5)^2 + X_1 \leq 0.3$ and $X_3^2 \leq X_4$ \\ \hline
\end{tabular}
\label{8treatment-decisionboundary}
}
\end{table}

Increasing the number of treatment arms complicates the classification problem, as a randomized treatment rule with equal probability yields a misclassification rate of 0.875. Figure \ref{fig:continouscorrectps-k8} demonstrates that the performance of all the methods becomes worse, where the proposed methods with covariate matching yield the best performance in all scenarios, including the settings with linear decision boundary. The unsatisfactory performance of Q-learning could be partly explained by the smaller sample size in each treatment arm, which could deteriorate the model fitting. Another possible explanation is that more parameters need to be estimated in Q-learning, as the treatment variable $A$ is encoded as $k-1$ dummy variables. These results highlight the superiority of multicategory M-learning with covariate matching.

Third, we adapt the setup in Section 4.1 to the scenarios with $p=18$. Specifically, the 18-dimensional covariates ${X}$ are independently generated from $U(0,1)$. The treatment $A$ conditioning on ${X}$ is simulated from a multinomial distribution taking values in $\{1,2,3,4\}$ with  $Pr(A = w \mid {X})=\frac{\exp ({X}^\top{\tilde{\beta}}_{w})}{\sum_{w^{\prime}=1}^{4} \exp ({X}^\top {\tilde{\beta}}_{w^{\prime}})}, \  w=1,2,3,4$.
Here, we set ${\tilde{\beta}}_{w}^\top=({\beta}_{w}^\top,{\beta}_{w}^\top,{\beta}_{w}^\top)$, where ${\beta}_{w}$ follows the same setup as in Section 4.1. Figure \ref{fig:continouscorrectps-p18} shows that Q-learning generally performs the best, with its advantage becoming pronounced in scenarios with linear decision boundary. Multi-AOL also exhibits competitive performance. Compared with the results in Section 4.1, the proposed methods with covariate matching deteriorate and yield poor performance in all the scenarios due to the matching on relatively high dimensional covariates. The additional 12 covariates are instruments that only relate to the treatment $A$, hence matching directly on all the covariates leads to unsatisfactory match sets.

We note that the proposed methods using generalized propensity scores matching still perform worse than those using covariate matching. Therefore, it is crucial to explore the scenario where multicategory M-learning based on generalized propensity scores may yield desirable performance.
In the real data analysis, we have found that LT is the optimal treatment for most patients, and the variants with  generalized propensity scores matching yield the best results. These observations motivate  us to investigate the setting with the presence of dominant class. We consider a new simulation setup with 12-dimensional covariates ${X}$ independently generated from $U(0,1)$. The treatment $A$ conditioning on ${X}$ is simulated from a multinomial distribution taking values in $\{1,2,3,4\}$ with $Pr(A = w \mid {X})=\frac{\exp ({X}^\top{\tilde{\beta}}^{*}_{w})}{\sum_{w^{\prime}=1}^{4} \exp ({X}^\top {\tilde{\beta}}^{*}_{w^{\prime}})}, \  w=1,2,3,4$. Here, we set ${\tilde{\beta^{*}_{w}}}^\top=({\beta}_{w}^\top,{\beta}_{w}^\top)$, where ${\beta}_{w}$ follows the same setup as in Section 4.1. The outcome variable $R$ is generated by $R=\sum_{w=1}^k I(w=A)(X^\top \gamma_w)+\epsilon$, where $\gamma_1=1.5\times(1,2,1,5,1,2,1,2,1,5,1,2)$, $\gamma_2=1.35\times(2,3,1,2,2,2,2,3,1,2,2,2)$, $\gamma_3=(3,1,2,1,1,4,3,1,2,1,1,4)$, $\gamma_4=(4,1,2,1,3,1,4,1,2,1,3,1)$, and $\epsilon$ is noisy variable generated independently from $U(0,1)$. In this simulation setting, treatment 1 is the optimal treatment  for most of the population. Additionally, the number of individuals whose optimal rule is treatment 2 is significantly larger than the number of  those whose optimal rule is treatment 3 or 4.

Figure \ref{fig:GPSmatchComparable-n1000p12} shows that Q-learning and Multi-AOL yield the best performance due to the correct model specification. In contrast, Multi-OL and AD-learning exhibit poor results with large variability, which can be explained by the small or extreme estimated propensity scores. Additionally,  the unnecessary group penalty could deteriorate the performance of AD-learning. Notably, the proposed methods with generalized propensity scores matching outperform the variants based on covariate matching, with larger value functions, lower misclassification rates, and smaller variability. This finding suggests that the proposed method with generalized propensity scores matching could be better suited for the scenarios with dominant treatment arms.
}

{\color{black}
\section{Additional Weighting Function}
As suggested by one reviewer, we  investigate an alternative weighting function:  the difference between the largest outcome and the second largest outcome, denoted by the suffix -gw3. To clearly demonstrate the performance distinctions, we present the misclassification rates for the variants of multicategory M-learning across nine scenarios.  Scenarios 1-4 are the settings LS, NS, LC, and NC reported in Figure 1, respectively. Scenarios 5-8 are the settings LS, NS, LC, and NC reported in Figure \ref{fig:survivalcoxph2}, respectively. Scenario 9 is the setting presented in Figure \ref{fig:GPSmatchComparable-n1000p12}.

Table \ref{g3_table} shows that the performances of different weighting functions are comparable. In Scenarios 1-4 with continuous outcomes, the variant -gw3-cov  slightly outperforms the other variants. However, in the remaining scenarios, the variants -gw1-cov, -gw1-gps, and -gw2-cov  yield slightly better performance. These results suggest that the performance of different weighting functions may vary across different scenarios.

\begin{table}[htbp]
\color{black}
    \centering
        \caption{Misclassification rate for  the variants of multicategory M-learning, averaged over 400 replications. The numbers in parentheses are standard deviations. The best results of value function are in bold. }
        %\vspace{1em}
        \resizebox{\textwidth}{!}{
             \begin{tabular}{ccccccccc}
             \toprule
              & \multicolumn{8}{c}{Variants of multicategory M-learning} \\
             \cmidrule{2-9}
              Scenarios & -g1-cov & -g1-gps & -gw1-cov &-gw1-gps & -gw2-cov & -gw2-gps & -gw3-cov & -gw3-gps  \\
            \midrule
            1   &  0.106 (0.021)  & 0.317 (0.069)  & 0.105 (0.021)  & 0.325 (0.07)  & 0.114 (0.024)  & 0.349 (0.072)  & \textbf{0.093} (0.017)  & 0.287 (0.062) \\
            2    &  0.308 (0.022)  & 0.526 (0.043)  & 0.281 (0.021)  & 0.504 (0.04)  & 0.288 (0.02)  & 0.514 (0.04)  & \textbf{0.277} (0.02)  & 0.503 (0.042)  \\
            3     &  0.104 (0.02)  & 0.326 (0.071)  & 0.101 (0.019)  & 0.364 (0.067)  & 0.104 (0.022)  & 0.379 (0.067)  & \textbf{0.097} (0.018)  & 0.329 (0.069) \\
            4     &  0.308 (0.022)  & 0.524 (0.044)  & 0.286 (0.02)  & 0.55 (0.052)  & 0.294 (0.021)  & 0.559 (0.053)  & \textbf{0.28} (0.019)  & 0.533 (0.049)\\
            5     &  0.139 (0.033)  & 0.396 (0.071)  & \textbf{0.138} (0.032)  & 0.362 (0.066)  & \textbf{0.138} (0.031)  & 0.367 (0.068) & 0.14 (0.029)  & 0.371 (0.065) \\
            6     &  0.353 (0.032)  & 0.589 (0.047)  & 0.336 (0.031)  & 0.569 (0.047)  & 0.337 (0.028)  & 0.57 (0.048) & \textbf{0.333} (0.031)  & 0.577 (0.047)  \\
            7     &  0.161 (0.037)  & 0.416 (0.072)  & \textbf{0.158} (0.033)  & 0.392 (0.072)  & 0.16 (0.036)  & 0.396 (0.074)  & 0.164 (0.034)  & 0.4 (0.067)  \\
            8      &  0.371 (0.036)  & 0.594 (0.048)  & \textbf{0.354} (0.035)  & 0.575 (0.046)  & 0.357 (0.031)  & 0.575 (0.049)  & 0.359 (0.037)  & 0.591 (0.048)  \\
            9     & 0.208 (0.07)  & 0.167 (0.052)  & 0.209 (0.073)  & \textbf{0.166} (0.061)  & 0.211 (0.068)  & 0.178 (0.056)  & 0.202 (0.074)  & 0.171 (0.058)  \\
            \bottomrule
             \end{tabular}
             }
             \label{g3_table}
\end{table}

}

\newpage
\section{Proof of Theorem 1}\label{proofthm1}
\begin{proof}
	For ease of exposition, we only discuss the situation with $k=3$ treatments. Extension to the setting with arbitrary number of treatments is straightforward. 
	
	For any subject $i$, there are two matched sets $\mathcal{M}^{(w)}_i$ with $w\neq A_i$. Let $w^{*}$ and $w^{**}$ be the other two treatments that are different from $A_i$. Select subjects $j$ and $k$ randomly from  these two matched sets $\mathcal{M}^{(w^{*})}_i$ and $\mathcal{M}^{(w^{**})}_i$. Then, $(A_i, A_j,A_k)=(A_i,w^{*},w^{**})$ is a permutation of $(1,2,3)$.  If $A_i=1$, then $A_j=w^{*}=2$ and $A_k=w^{**}=3$; if $A_i=2$, then $A_j=w^{*}=3$ and $A_k=w^{**}=1$; if $A_i=3$, then $A_j=w^{*}=1$ and $A_k=w^{**}=2$. 
	
	The multicategory matching-based value function can be written as
	\begin{align*}
		&V_n(\mathcal{D};g)\\
		=&\frac{1}{n}\sum_{i=1}^{n}\frac{1}{|\mathcal{M}^{(w^{*})}_i| |\mathcal{M}^{(w^{**})}_i|}\sum_{j \in \mathcal{M}^{(w^{*})}_i}\sum_{k \in \mathcal{M}^{(w^{**})}_i} I\{ \mathcal{D}(X_i)=h(R_i,R_j,R_k,A_i,A_j,A_k)\} g(R_i,R_j,R_k),
	\end{align*}
	where 
	\begin{align*}
		&h(R_i,R_j,R_k,A_i,A_j,A_k)\\
		=&I(R_i \geq R_j,R_i \geq R_k)A_i+I(R_j \geq R_i,R_j \geq R_k)A_j+ I(R_k \geq R_i,R_k \geq R_j)A_k
	\end{align*}
	is the treatment corresponding to the largest value among $\{R_i,R_j,R_k\}$.
	
	Let $(\widetilde{X},\widetilde{A},\widetilde{R})$ and $(\overline{X},\overline{A},\overline{R})$ be two i.i.d. copies of $(X,A,R)$. In addition, let E, $\widetilde{E}$, and $\overline{E}$ denote the expectation w.r.t. $(X,A,R)$, $(\widetilde{X},\widetilde{A},\widetilde{R})$, and $(\overline{X},\overline{A},\overline{R})$ respectively. Let $w^{*}(A)$ and  $w^{**}(A)$ be the other two treatments that are different 
	from $A$. That is, $(A,w^{*}(A),w^{**}(A))$ formulates a permutation of $(1,2,3)$.
	
	After some algebra, the asymptotic limit of $V_n(\mathcal{D};g)$ is
	\begin{equation*}
		\begin{aligned}
			&V(\mathcal{D};g)\\
			=&E\left[\widetilde{E}\left\{ \overline{E}\left[I\{ \mathcal{D}(X)=h(R,\widetilde{R},\overline{R},A,\widetilde{A},\overline{A})\} g(R,\widetilde{R},\overline{R}) \mid \overline{A}=w^{**}(A), \overline{X}=X \right] \mid \widetilde{A}=w^{*}(A), \widetilde{X}=X \right\}\right] \\
			=&E\left[\widetilde{E}\left\{ \overline{E}\left[I\{ \mathcal{D}(X)=h(R,\widetilde{R},\overline{R},A,\widetilde{A},\overline{A})\} g(R,\widetilde{R},\overline{R}) \mid \overline{A}=w^{**}(A), \widetilde{A}=w^{*}(A), \widetilde{X}=\overline{X}=X \right] \right\}\right].
		\end{aligned}
	\end{equation*}
	
	By the law of iterated expectation, we note the following expression:
	\begin{equation*}
		V(\mathcal{D};g)=E_X \left[I\{\mathcal{D}(X)=1\}\times E_1+I\{\mathcal{D}(X)=2\}\times E_2+I\{\mathcal{D}(X)=3\}\times E_3\right],
	\end{equation*}
	where 
	\begin{equation*}
		\begin{aligned}
			&E_1\\
			=&P(A=1|X)E\left[I\{h(R,\widetilde{R},\overline{R},1,2,3)=1\} g(R,\widetilde{R},\overline{R})\mid A=1,\widetilde{A}=2,\overline{A}=3,\widetilde{X}=\overline{X}=X\right] \\ &+P(A=2|X)E\left[I\{h(R,\widetilde{R},\overline{R},2,3,1)=1\} g(R,\widetilde{R},\overline{R})\mid A=2,\widetilde{A}=3,\overline{A}=1,\widetilde{X}=\overline{X}=X\right] \\
			&+P(A=3|X)E\left[I\{h(R,\widetilde{R},\overline{R},3,1,2)=1\} g(R,\widetilde{R},\overline{R})\mid A=3,\widetilde{A}=1,\overline{A}=2,\widetilde{X}=\overline{X}=X\right] \\
			=&P(A=1|X) E\left[I(R\geq \widetilde{R},R \geq \overline{R}) g(R,\widetilde{R},\overline{R})\mid A=1,\widetilde{A}=2,\overline{A}=3,\widetilde{X}=\overline{X}=X\right] \\
			&+ P(A=2|X)E\left[I(\overline{R} \geq R,\overline{R} \geq \widetilde{R}) g(R,\widetilde{R},\overline{R})\mid A=2,\widetilde{A}=3,\overline{A}=1,\widetilde{X}=\overline{X}=X\right] \\
			&+ P(A=3|X)E\left[I(\widetilde{R} \geq R,\widetilde{R} \geq \overline{R}) g(R,\widetilde{R},\overline{R})\mid A=3,\widetilde{A}=1,\overline{A}=2,\widetilde{X}=\overline{X}=X\right]\\
			=&E\left[I(R\geq \widetilde{R},R \geq \overline{R}) g(R,\widetilde{R},\overline{R})\mid A=1,\widetilde{A}=2,\overline{A}=3,\widetilde{X}=\overline{X}=X\right].
		\end{aligned}
	\end{equation*}
	Here, the last equation is due to the exchangeability of $g(\cdot)$.
	For $k=3$, 
	\[\Delta_g(r_1,r_2,w,X)=E\left[I\{R \geq r_1, R \geq r_2\} \times g(R,r_1,r_2) \mid A=w, X\right], \text{ for } w=1,2,3.
	\]
	Therefore,
	\begin{equation*}
		E_1=\int \Delta_g(r_1,r_2,1,X) \mathrm{d} F(r_1,r_2 \mid a_1=2,a_2=3, X),
	\end{equation*}
	where $ F(r_1,r_2 \mid a_1=2,a_2=3, X)$ is the conditional distribution of $(\widetilde{R},\overline{R})$ given $(\widetilde{A},\overline{A})=(2,3)$ and $X$.
	Similarly, we have
	\begin{equation*}
		E_2=\int \Delta_g(r_1,r_2,2,X) \mathrm{d} F(r_1,r_2 \mid a_1=1,a_2=3, X).
	\end{equation*}
	and
	\begin{equation*}
		E_3=\int \Delta_g(r_1,r_2,3,X) \mathrm{d} F(r_1,r_2 \mid a_1=1,a_2=2, X).
	\end{equation*}
	
	Now, it is clear that the optimal  rule   $\mathcal{D}^*(x)=\underset{k \in \{1,2,3\}}{\arg\max} \ E_k$
	maximizes the limiting value function $V(\mathcal{D};g)$ given  covariate $X=x$ and the weighting function $g(\cdot)$.
\end{proof}

\newpage
{\color{black}
\section{Additional Figures}
}

\begin{figure}[H]
	\centering
	\includegraphics[width=1\linewidth]{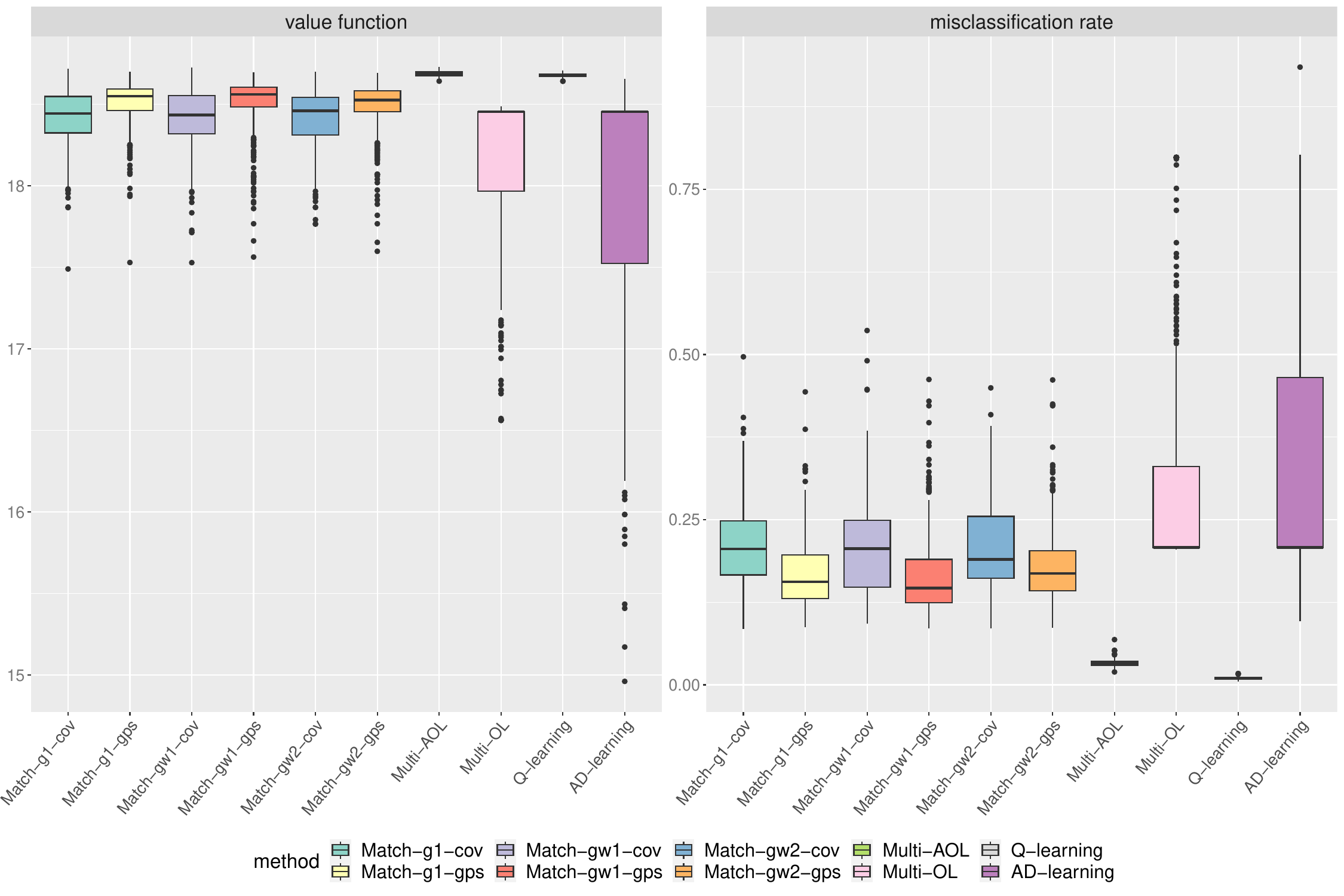}
	\caption{Boxplots for empirical value function and misclassification rate in the presence of continuous outcomes.  The simulation setup is adapted from Section 4.1,  with the covariate dimension $p$ increased to 12. The outcome variable is generated by $R=\sum_{w=1}^k I(w=A)(X^\top \gamma_w)+U(0,1)$. }
	\label{fig:GPSmatchComparable-n1000p12}
\end{figure}

\newpage

\bibliographystyle{ecta} 
\small{
\bibliography{revised-supplemental-paper-ref.bib}
}